\documentclass[
  journal=pasa,
  manuscript=research-paper,
  year=2022,
]{cup-journal}

\usepackage{microtype,siunitx,booktabs,amssymb,amsmath,xcolor,multirow,longtable,pdflscape,threeparttablex,bm}
\usepackage[mathlines]{lineno}  
\definecolor{link}{rgb}{0,0,1}
\usepackage[pdfnewwindow=true,%
            colorlinks=true,
            linkcolor=link,
            citecolor=link,
            filecolor=link,
            urlcolor=link, 
            ]{hyperref}
\usepackage{xurl}
\newcommand{\om}{\ensuremath{\Omega_{\text{m}}}}

\newcommand{\Pippin}{\textsc{Pippin}}
\newcommand{\SNANA}{\texttt{SNANA}}

\newcommand{\zCMB}[1]{\ensuremath{z_{\text{CMB}}^{\mathrm{#1}}}}
\newcommand{\zHD}[1]{\ensuremath{z_{\text{HD}}^{\mathrm{#1}}}}

\newcommand{\zSun}[1]{\ensuremath{z_{\text{Sun}}^{\mathrm{#1}}}}
\newcommand\zhel{\ensuremath{z_{\text{hel}}}}
\newcommand{\znew}{$z_{\text{new}}$}
\newcommand{\zold}{$z_{\text{old}}$}
\newcommand\dg{\ensuremath{^{\circ}}}
\newcommand\vSun[1]{\ensuremath{v_{\text{Sun}}^{\mathrm{#1}}}}
\newcommand\voC{\ensuremath{v_{\text{Sun}}^{\text{COBE}}}}
\newcommand\voP{\ensuremath{v_{\text{Sun}}^{\text{Planck}}}}

\newcommand{\Vext}{\ensuremath{\bm{V}_{\text{ext}}}}
\newcommand{\vp}{\ensuremath{v_{\text{p}}}}

\newcommand{\zp}{\ensuremath{z_{\text{p}}}}
\newcommand{\ten}[1]{\ensuremath{10^{#1}}}
\newcommand{\tten}[1]{\ensuremath{\times 10^{#1}}}
\newcommand{\kmsMpc}{km\,s$^{-1}$Mpc$^{-1}$}
\newcommand{\hMpc}{$h^{-1}$Mpc}
\newcommand{\kms}{km\,s$^{-1}$}
\newcommand{\Hi}{H\,\textsc{i}}
\newcommand{\papers}{\url{https://PantheonPlusSH0ES.github.io}}
\newcommand{\data}{\url{https://github.com/PantheonPlusSH0ES/DataRelease}}
\newcommand{\pvcode}{\url{https://github.com/KSaid-1/pvhub}}

\title{The Pantheon+ Analysis: Improving the Redshifts and Peculiar Velocities of Type Ia Supernovae Used in Cosmological Analyses}

\author{Anthony Carr}
\affiliation{School of Mathematics and Physics, University of Queensland, Brisbane, QLD 4072, Australia}
\email[Anthony Carr]{anthony.carr@uq.net.au}

\author{Tamara M.\ Davis}
\affiliation{School of Mathematics and Physics, University of Queensland, Brisbane, QLD 4072, Australia}

\author{Dan Scolnic}
\affiliation{Department of Physics, Duke University, Durham, NC 27708, USA}

\author{Khaled Said}
\affiliation{School of Mathematics and Physics, University of Queensland, Brisbane, QLD 4072, Australia}

\author{Dillon Brout}
\affiliation{Center for Astrophysics, Harvard \& Smithsonian, 60 Garden Street, Cambridge, MA 02138, USA}
\alsoaffiliation{NASA Einstein Fellow}

\author{Erik R.\ Peterson}
\affiliation{Department of Physics, Duke University, Durham, NC 27708, USA}

\author{Richard Kessler}
\affiliation{Kavli Institute for Cosmological Physics, University of Chicago, Chicago, IL 60637, USA}
\alsoaffiliation{Department of Astronomy and Astrophysics, University of Chicago, Chicago, IL 60637, USA}

\doi{10.1017/pasa.2022.41}

\received {26 05 2022}
\revised  {05 08 2022}
\accepted {29 08 2022}
\published{11 10 2022}


\begin{document}

\begin{abstract}
We examine the redshifts of a comprehensive set of published Type Ia supernovae, and provide a combined, improved catalogue with updated redshifts.
We improve on the original catalogues by using the most up-to-date heliocentric redshift data available; ensuring all redshifts have uncertainty estimates; using the exact formulae to convert heliocentric redshifts into the Cosmic Microwave Background (CMB) frame; and utilising an improved peculiar velocity model that calculates local motions in redshift-space and more realistically accounts for the external bulk flow at high-redshifts.
We review 2607 supernova redshifts; 2285 are from unique supernovae and 322 are from repeat-observations of the same supernova. 
In total, we updated 990 unique heliocentric redshifts, and found 5 cases of missing or incorrect heliocentric corrections, 44 incorrect or missing supernova coordinates, 230 missing heliocentric or CMB frame redshifts, and 1200 missing redshift uncertainties.
The absolute corrections range between $10^{-8} \leq \Delta z \leq 0.038$, and RMS$(\Delta z) \sim 3\tten{-3}$.  
The sign of the correction was essentially random, so the  mean and median corrections are small: 4\tten{-4} and 4\tten{-6} respectively.  
We examine the impact of these improvements for $H_0$ and the dark energy equation of state $w$ and find that the cosmological results change by $\Delta H_0 = -0.12$ \kmsMpc\ and $\Delta w = 0.003$, both significantly smaller than previously reported uncertainties for $H_0$ of 1.0 \kmsMpc\ and $w$ of 0.04 respectively.
\end{abstract}

\section{Introduction} \label{sec:intro}

The power of Type Ia Supernovae (SNe Ia) as a probe of the expansion history of the universe comes from comparing the measured distances of the SNe to the distances expected for their redshift in different cosmological models \citep{Riess1998,Perlmutter1999,WoodVasey2007,Kessler2009,Betoule2014,Scolnic2018,DES2019}. 
Since the relative precision of spectroscopically measured redshifts is typically significantly greater than that of redshift-independent distances, much more effort has been spent on improving distance measurements than improving redshift measurements \citep[e.g.][]{Phillips1993, Phillips1999, Goldhaber2001, Guy2007, Jha2007, Hicken2009, Kessler2009, Scolnic2015, Kessler2017, Brout2019, Kessler2019, Lasker2019}.  
This prioritisation is supported by the fact that redshift measurements, either from the host galaxies or SNe, are straightforward; small errors are usually expected to be random, shifting redshifts higher as often as lower. 
However, with samples of greater than 1000 SNe, systematic uncertainties are of paramount concern, and potential systematic biases in the redshift measurements must be considered \citep[e.g.][]{Huterer2004,Wojtak2015,Davis2019,Steinhardt2020,Mitra2021}.  
In this analysis, we perform a comprehensive review of the redshifts of individual SNe used in the latest samples for cosmological analyses and analyse potential biases due to issues with redshifts in the recovery of cosmological parameters.

Several recent papers have considered the impact of redshifts errors on supernova cosmology.
For example, \citet{Steinhardt2020} determined whether the source for each redshift in the Pantheon sample \citep{Scolnic2018} was either the host galaxy spectrum or SN spectrum, and found difference in cosmological parameters at a $\sim 3\sigma$ level between the two subsets. 
\citet{Rameez2021} noted changes in the measured redshifts of sub-samples of large SN compilations that were larger than the reported uncertainties and questioned the repeatability of SN experiments.  
While here we show the effect of redshift errors on cosmological parameters remains small (relative to their current precision), we note that the redshifts came from a variety of sources, with many measurements having been over 20 years ago. 
Old and/or inhomogeneous redshift measurements are not necessarily a problem, but these factors increase the potential for miscellaneous errors to be carried through different SN samples, so a comprehensive review is warranted.
Achieving accurate redshifts for cosmological studies requires multiple stages, and in this paper we apply improvements at each stage except for performing new spectroscopic measurements.

Redshifts in the heliocentric frame are measured either from the SN spectrum, which is typically precise on the level of $\sigma_z\sim0.005$ (a somewhat conservative estimation), or the host-galaxy spectrum, which is typically precise on the level of $\sigma_z\sim0.0002$ (see Section \ref{sec:uncertainties}).
Host redshifts are preferred because the hosts have sharper spectral lines that result in a more accurate and precise redshift. 
However, it is essential for the correct host-galaxy to be associated with the SN \citep{Gupta2016}, else SNe will be misplaced on the Hubble diagram.  
Here we review the host galaxy assignment of all SNe where possible, and update heliocentric redshifts accordingly.   

Once the heliocentric redshift is determined, we convert the redshift into the CMB frame.  
While the CMB conversion is standard, an unnecessary approximation has often been applied \citep[see, e.g.][and references therein]{Carr2021} and we replace that with the exact correction (Section \ref{sec:convert2cmb}).  

The final step to obtain accurate cosmological redshifts is applying the correction to account for the peculiar velocity of the source.  
We introduce a slightly improved technique of estimating peculiar velocities (that also better models the external bulk flow to arbitrary redshift) based on the existing 2M++ compilation \citep{Carrick2015}.
We apply this correction to all redshifts whereas previously, corrections had been applied only at low redshift, or with a biased model at large redshifts.

We thus release a comprehensive update to the redshifts of all publicly available Type Ia supernovae that make up the `Pantheon+' sample.  
Unlike previous analyses, we do not isolate our work to redshifts of `cosmologically useful' SNe (those that make it onto the Hubble diagram) since data cuts may be relaxed or otherwise altered in future analyses.  

This paper is one of many that contribute to the Pantheon+ sample, culminating in the full cosmology analysis in \citet{Brout2022cosmo}.
This work is companion to \citet{Peterson2021}, that studies the effects of replacing host galaxy redshifts with average redshifts of host galaxy groups on Hubble diagram residuals, and provides group-averaged redshifts and group-centre coordinates which we also release here.
In addition, \citet{Peterson2021} studies the efficacy of using different peculiar velocity samples---including those derived in this work---on Hubble residuals.
\citet{Brout2022cals} re-calibrates the many photometric systems of Pantheon+ and quantifies the systematic effects of photometric calibration on cosmological parameters.
\citet{Scolnic2022samples} releases the 1701 updated light curves of 1550 unique SNe Ia\footnote{The cosmology sample of this work differs in number to the full analysis due to our simplified data cuts and lack of full covariance analysis.} used in the Pantheon+ $w$ analysis and Supernovae and H0 for the Equation of State of dark energy (SH0ES) $H_0$ analysis \citep{Riess2022}.
\citet{Peterson2021}, \citet{Scolnic2022samples} and \cite{Brout2022cals, Brout2022cosmo} utilise the redshifts of this work, and we utilise the group-averaged redshifts from \citet{Peterson2021} and distance moduli from \citet{Brout2022cosmo} in our analysis of the effects of the redshift updates.
See \papers{} for the other papers that contribute to Pantheon+.

In this work, the main improvements we implemented are detailed in the following sections: 
\begin{itemize}
    \item Fixed coordinates and miscellaneous bookkeeping redshift errors (Section \ref{sec:sample_description}).
    \item Updated heliocentric redshift values using the NASA/IPAC Extragalactic Database (NED) when better redshifts were available (Section \ref{sec:validhelio}).
    \item Ensured all redshifts have uncertainty estimates (Section \ref{sec:uncertainties}).
    \item Used the exact redshift conversion when (a) going from heliocentric redshifts to the CMB frame, and (b) going from CMB frame to Hubble diagram redshift (Section \ref{sec:convert2cmb}).
These respectively correct for (a) our Sun's motion with respect to the CMB and (b) the host galaxy's motion with respect to the CMB.
    \item Provided improved peculiar velocity estimates that better represent the bulk flow at large distances (Section \ref{sec:pv}).
\end{itemize}

Next, we analyse the impact of each change on cosmological parameters in Section \ref{sec:impact} and finally discuss and conclude in Section \ref{sec:discussion_conclusion}.

The Pantheon+ redshifts and accompanying data will be released as a machine readable Centre de Donn\'{e}es astronomiques de Strasbourg (CDS) VizieR table with the publication of this work, and also on GitHub at \data{} which will log any possible updates.
The light curves for each SN \citep[see][]{Scolnic2022samples}, which contain our updated redshifts and peculiar velocities, are also available at this GitHub.
The peculiar velocity method developed for this paper is available at \pvcode{}.

\section{Samples and bookkeeping}\label{sec:sample_description}
\begin{figure*}[ht]
    \centering
    \includegraphics[width=0.90\textwidth]{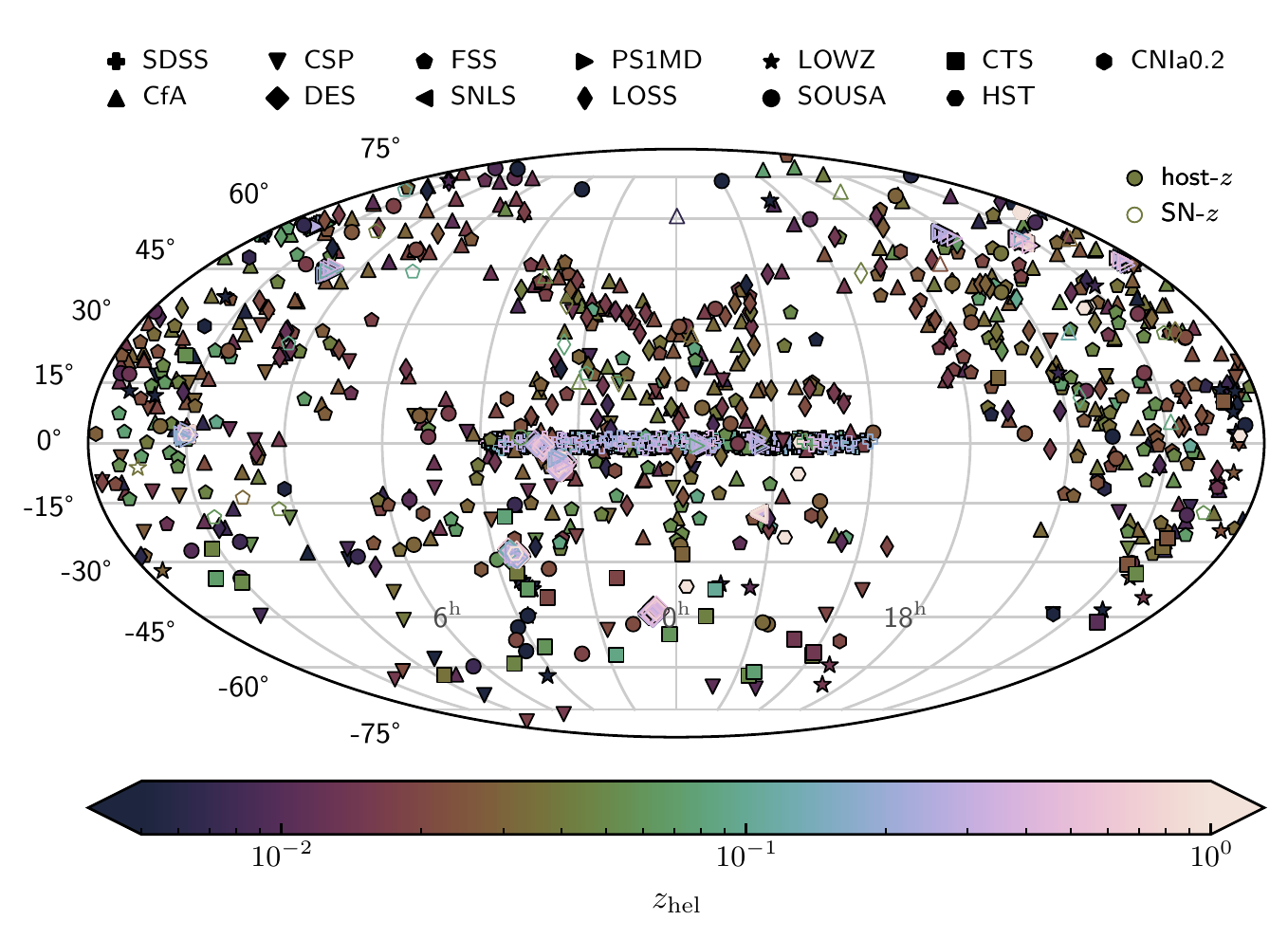}
    \caption{Distribution of Pantheon+ SNe across the sky. Many SNe are common between samples, but only one sample is picked to represent each SN. Redshifts from the host galaxies are represented by black-outlined solid symbols, and redshifts from supernova spectra are colour-outlined unfilled symbols. Several samples (generally at higher $z$ i.e.~SNLS, DES, PS1MD, HST) targeted small sky areas repeatedly, so many SNe are confined to small patches. These patches are visible where many lighter, unfilled symbols overlap (these groups obscure the underlying host galaxy redshifts which are always the majority).  
    }
    \label{fig:skyplot}
\end{figure*}

Our aim is to complete a comprehensive review of the redshifts assigned to every publicly available SN Ia used for cosmology and other SN Ia studies.  
We include primarily normal Type Ia supernovae along with various Ia subtypes, such as `1991T--like' or just `peculiar' \citep{Li2001}. 
The Pantheon+ sample is compiled of supernovae taken from a diverse array of samples, as listed in Table~\ref{tab:panthplus_description} and shown in Figure~\ref{fig:skyplot}.
The master list of our updated redshifts, which includes the SN and host coordinates; heliocentric, CMB, and cosmological (Hubble diagram) redshifts; and peculiar velocity values, can be found in Table \ref{tab:master}.  
Including all of these quantities aids in the traceability of the redshifts (and hosts) and repeatability of the corrections.

\begin{table*}
\renewcommand{\arraystretch}{0.95}
\begin{threeparttable}[hbt!]
\caption{Description of sub-samples and list of changes. $N_{\text{Tot}}$ is the number of Type Ia (including subtypes) supernova light curves in each sample. See Section \ref{sec:validhelio} for the definition of `best NED redshift'. See dot points in Section~\ref{subsec:corrections} and Section~\ref{sec:validhelio} for more detailed descriptions of the improvements. \label{tab:panthplus_description}}
\begin{tabular}{p{5cm}p{1.2cm}p{0.9cm}lcp{6cm}}
\toprule
Source & Abbrev. &  Ref. & $N_{\text{Tot}}$ & \zhel{} range & Improvements \\
\midrule
Hubble Deep Field North (using HST) & HDFN & \centering 1, 2 & 1 & 1.755 & Corrected \zhel{} being listed as \zCMB{}.\\
Supernova Cosmology Project (using HST) & SCP & \centering 3 & 8 & 1.014--1.415 & Corrected \zhel{} being listed as \zCMB{}. Added coordinates. Reassigned uncertainties based on host or SN redshift.\\
Cosmic Assembly Near Infra-Red Deep Extragalactic Legacy Survey and Cluster Lensing And Supernova survey with Hubble (using HST) & CANDELS +CLASH & \centering 4 & 13 & 1.03--2.26 & Corrected \zhel{} being listed as \zCMB{}. Updated several \zhel{} to originally published values. Updated name and coordinates of one SN.\tnote{a}\\
Complete Nearby (Redshift $<0.02$) Sample of Type Ia Supernova Light Curves & CNIa0.02 & \centering 5 & 17 & 0.0041--0.0303 & Found best NED \zhel{} or added uncertainties where necessary. \\
Center for Astrophysics (1) & CfA1 & \centering 6 & 22 & 0.0031--0.123 & Corrected coordinates of one SN. Found best NED \zhel{}. Added uncertainties. Identified hosts.\\
Cal\'{a}n/Tololo survey & CTS & \centering 7 & 29 & 0.0104--0.101 & Found best NED \zhel{}. Added uncertainties. Identified hosts.\\
Great Observatories Origins Deep Survey and Probing Acceleration Now with Supernova (using HST) & GOODS +PANS & \centering 8, 9 & 29 & 0.457--1.390 & Corrected \zhel{} being listed as \zCMB{}. Added coordinates. Reassigned uncertainties based on host or SN redshift.\\
Center for Astrophysics (2) & CfA2 & \centering 10 & 44 & 0.0067--0.0542 & Found best NED \zhel{}. Added uncertainties. Identified hosts.\\
Low-redshift (various sources) & LOWZ & \centering 11--23 & 66 & 0.0014--0.038 & Found best NED \zhel{}.  Added uncertainties. Identified hosts.\\
Lick Observatory Supernova Search (2005--2018) & LOSS2 & \centering 24 & 78 & 0.0008--0.082 & Found best NED \zhel{}. Added uncertainties where necessary. Identified hosts.\\
Center for Astrophysics (4p1, 4p2) & CfA4 & \centering 25 & 94 & 0.0067--0.0745 & Corrected coordinates of one SN. Found best NED \zhel{}.  Added uncertainties. Identified hosts.\\
Swift Optical/Ultraviolet Supernova Archive & SOUSA & \centering 26, 27 & 121 & 0.0008--0.0616 & Found best NED \zhel{}. Added uncertainties. Identified hosts.\\
Carnegie Supernova Project (DR3) & CSP & \centering 28 & 134 & 0.0038--0.0836 & Found best NED \zhel{}. Added uncertainties. Identified hosts.\\
Lick Observatory Supernova Search (1998--2008) & LOSS1 & \centering 29 & 165 & 0.0020--0.0948 & Found best NED \zhel{}. Added uncertainties. Identified hosts.\\
Center for Astrophysics (3S, 3K) & CfA3 & \centering 30 & 185 & 0.0032--0.084 & Found best NED \zhel{}. Added uncertainties. Identified hosts.\\
SuperNova Legacy Survey & SNLS & \centering 31 & 239 & 0.1245--1.06 & Identified 10 hosts and updated those redshifts. Added uncertainties.\\
Foundation Supernova Survey & FSS & \centering 32, 33 & 242 & 0.0045--0.1106 & Corrected coordinates of six SNe. Updated one redshift.\tnote{b}\\
Dark Energy Survey (3YR) & DES & \centering 34, 35 & 251\tnote{c} & 0.0176--0.850 & Updated redshifts and reassigned uncertainties based on host or SN redshift. One additional redshift update.\tnote{d}\\
Panoramic Survey Telescope \& Rapid Response System Medium Deep Survey & PS1MD & \centering 36 & 370 & 0.0252--0.670 & Identified 20 hosts and updated those redshifts. Updated one additional redshift.\tnote{e}\\
Sloan Digital Sky Survey & SDSS & \centering 37 & 499\tnote{f} & 0.0130--0.5540 &  Identified hosts. Updated 127 redshifts.\\
\bottomrule
\end{tabular}
\begin{tablenotes}[flushleft]
\item [a] SNID vespasion: Pantheon ID was previously `vespesian' (spelled with an {\em e}), and coordinates were mistakenly those of the supernova Obama (EGS11Oba) \citep{Riess2018}. See new coordinates in Table \ref{tab:FSSpos}.
\item [b] PSNJ1628383 from FSS shares the same host group as SN 2009eu. See entry in Table \ref{tab:update_outliers}.
\item [c] The DES sample contains 26 non-confirmed but highly probable SNe Ia spectroscopic classifications \citep[type `SNIa?' in Table 6 of][]{Smith2020}. Of these, 22 pass our cosmology sample quality cuts.
\item [d] Updated with new OzDES redshifts and improved SN host association (M.~Smith, private communication).  The redshift of DES supernova 1280201 (DES15X3iv) was updated to the higher precision FSS measurement (ASASSN-15od).
\item [e] The PS1MD redshift of 580104 was updated to the higher precision DES measurement (1261579; DES13X3woy).
\item [f] We include only secure spectroscopic Type Ia classifications from SDSS, which excludes 41 SNe \citep[type `SNIa?', in Table 2 of][]{Sako2018}, four of which were originally in Pantheon (2005gv, 2005kt, 2007oq, 2007ow) cf.~22/26 DES `SNIa?' in our cosmology sample, implying the SDSS SNIa?~are less secure. A further two SNe (2006lo, 2006md) in Pantheon were excluded due to being photometrically classified \citep[type `zSNIa' in][]{Sako2018}.
\item \textbf{References}: (1) \citet{Gilliland1998}; (2) \citet{Riess2001}; 
(3) \citet{Suzuki2012}; 
(4) \citet{Riess2018}; 
(5) \citet{Chen2020}; 
(6) \citet{Riess1999}; 
(7) \citet{Hamuy1996}; 
(8) \citet{Riess2004}; (9) \citet{Riess2007}; 
(10) \citet{Jha2006}; 
(11) \citet{Jha2007} and references therein; (12) \citet{Milne2010}; (13) \citet{Stritzinger2010}; (14) \citet{Tsvetkov2010}; (15) \citet{Zhang2010}; (16) \citet{Hsiao2015}; (17) \citet{Krisciunas2017_2005df}; (18) \citet{Burns2018}; (19) \citet{Contreras2010}; (20) \citet{Gall2018}; (21) \citet{Wee2018}; (22) \citet{Burns2020}; (23) \citet{Kawabata2020}; 
(24) \citet{Stahl2019}; 
(25) \citet{Hicken2012}; 
(26) \citet{Brown2014}; (27) \url{https://pbrown801.github.io/SOUSA/}; 
(28) \citet{Krisciunas2017}; 
(29) \citet{Ganeshalingam2010}; 
(30) \citet{Hicken2009}; 
(31) \citet{Guy2010}; 
(32) \citet{Foley2018}; 
(33) \citet{Scolnic2022samples}; 
(34) \citet{Brout2019}; (35) \citet{Smith2020}; 
(36) \citet{Scolnic2018}; 
(37) \citet{Sako2018}. 
\end{tablenotes}
\end{threeparttable}
\end{table*}

\begin{table*}[h!]
\renewcommand{\arraystretch}{1.0}
\caption{Master table of updated redshifts. Symbols are defined in Section \ref{subsec:description}. The full machine readable table is available online from VizieR and at \data{}. The online versions of this table include columns for peculiar velocity uncertainty and binary classifications for if a SN has an associated host, if the redshift is from a host and if the supernova has group values. All peculiar velocities have an uncertainty of 250 \kms{}, and all \zhel{} share the same uncertainty as their corresponding \zCMB{}. Blank entries for group information mean the SN has no associated group, and blank IAUC entries mean there is no IAU name for the SN. \label{tab:master}} 
\begin{tabular}{lllS[table-format=3.6]S[table-format=2.6]S[table-format=3.6]S[table-format=2.6]llrc}
\toprule
 &  &  & \multicolumn{1}{c}{SN RA} & \multicolumn{1}{c}{SN Dec} & \multicolumn{1}{c}{Host RA} & \multicolumn{1}{c}{Host Dec} &  &  & & \\
\multicolumn{1}{c}{SNID} & \multicolumn{1}{c}{IAUC} & \multicolumn{1}{c}{Host} & \multicolumn{1}{c}{\dg{} (J2000)} & \multicolumn{1}{c}{\dg{} (J2000)} & \multicolumn{1}{c}{\dg{} (J2000)} & \multicolumn{1}{c}{\dg{} (J2000)} & \multicolumn{1}{c}{\zhel{}} & \multicolumn{1}{c}{\zCMB{}} & \multicolumn{1}{c}{$\sigma_{z_\text{hel}}$}  & \dots\\
\multicolumn{1}{c}{(1)} & \multicolumn{1}{c}{(2)} & \multicolumn{1}{c}{(3)} & \multicolumn{1}{c}{(4)} & \multicolumn{1}{c}{(5)} & \multicolumn{1}{c}{(6)} & \multicolumn{1}{c}{(7)} & \multicolumn{1}{c}{(8)} & \multicolumn{1}{c}{(9)} & \multicolumn{1}{c}{(10)} & \\
\midrule
2001G & 2001G & MCG +08-17-043 & 137.38825 & 50.28092 & 137.38721 & 50.28186 & 0.016703 & 0.017272 & 1.1\tten{-5}  & \dots\\
2001V & 2001V & NGC 3987 & 179.35388 & 25.20250 & 179.33717 & 25.19539 & 0.015007 & 0.016045 & 9.0\tten{-4}  & \dots\\
2001ah & 2001ah & UGC 06211 & 167.62425 & 55.16083 & 167.62646 & 55.16983 & 0.057763 & 0.058373 & 1.5\tten{-5}  & \dots\\
2001az & 2001az & UGC 10483 & 248.61546 & 76.02967 & 248.62017 & 76.02972 & 0.040695 & 0.040593 & 9.0\tten{-5}  & \dots\\
2001da & 2001da & NGC 7780 & 358.38658 & 8.11739 & 358.38404 & 8.11814 & 0.017381 & 0.016148 & 7\tten{-6}  & \dots\\
2001en & 2001en & NGC 0523 & 21.34542 & 34.02514 & 21.33637 & 34.02494 & 0.015881 & 0.014937 & 7\tten{-6}  & \dots\\
2001fe & 2001fe & UGC 05129 & 144.48792 & 25.49481 & 144.49167 & 25.49478 & 0.013514 & 0.014478 & 8\tten{-6}  & \dots\\
2001gb & 2001gb & IC 0582 & 149.75400 & 17.82011 & 149.75096 & 17.81714 & 0.025439 & 0.026529 & 1.1\tten{-5}  & \dots\\
2001ic & 2001ic & NGC 7503 & 347.68058 & 7.56956 & 347.67617 & 7.56769 & 0.044089 & 0.042802 & 9\tten{-6}  & \dots\\
2002bf & 2002bf & CGCG 266-031 & 153.92629 & 55.66853 & 153.92604 & 55.66747 & 0.024376 & 0.024936 & 7\tten{-6}  & \dots\\
\bottomrule
\end{tabular}
\end{table*}

\addtocounter{table}{-1}
\begin{table*}
\begin{threeparttable}[h!]
\caption{Master table of updated redshifts (continued).} 
\begin{tabular}{lclrcS[table-format=3.6]S[table-format=2.6]S[table-format=1.6]S[table-format=1.6]S[table-format=1.6]c}
\toprule
& & &  & \multicolumn{1}{c}{\vp{}} & \multicolumn{1}{c}{Group RA} & \multicolumn{1}{c}{Group Dec} &  &  &  & Group \vp{}\\
\multicolumn{1}{c}{SNID} & \dots & \multicolumn{1}{c}{\zHD{}} & \multicolumn{1}{c}{$\sigma_{z_\text{HD}}$} & \multicolumn{1}{c}{\kms{}} & \multicolumn{1}{c}{\dg{} (J2000)} & \multicolumn{1}{c}{\dg{} (J2000)} & \multicolumn{1}{c}{Group} \zhel{} & \multicolumn{1}{c}{Group \zCMB{}} & \multicolumn{1}{c}{Group \zHD{}} & \multicolumn{1}{c}{\kms{}}\\
\multicolumn{1}{c}{(1)} & & \multicolumn{1}{c}{(11)} & \multicolumn{1}{c}{(12)} & \multicolumn{1}{c}{(13)} & \multicolumn{1}{c}{(14)} & \multicolumn{1}{c}{(15)} & \multicolumn{1}{c}{(16)} & \multicolumn{1}{c}{(17)} & \multicolumn{1}{c}{(18)} & \multicolumn{1}{c}{(19)} \\
\midrule
2001G & \dots & 0.01770 & 8.5\tten{-4} & -127 & {\dots} & {\dots} & \dots & \dots & \dots & \dots\\
2001V & \dots & 0.01570 & 1.2\tten{-3} & 102 & 179.514560 & 25.171888 & 0.014883 & 0.015920 & 0.01557 & 102\\
2001ah & \dots & 0.05886 & 8.8\tten{-4} & -138 & {\dots} & {\dots} & \dots & \dots & \dots & \dots\\
2001az & \dots & 0.04101 & 8.7\tten{-4} & -121 & {\dots} & {\dots} & \dots & \dots & \dots & \dots\\
2001da & \dots & 0.01659 & 8.5\tten{-4} & -129 & 358.404510 & 7.930666 & 0.017816 & 0.016583 & 0.01702 & -176\\
2001en & \dots & 0.01509 & 8.5\tten{-4} & -46 & 21.041921 & 33.581505 & 0.016294 & 0.015343 & 0.01550 & -46\\
2001fe & \dots & 0.01467 & 8.5\tten{-4} & -57 & {\dots} & {\dots} & \dots & \dots & \dots & \dots\\
2001gb & \dots & 0.02650 & 8.6\tten{-4} & 9 & {\dots} & {\dots} & \dots & \dots & \dots & \dots\\
2001ic & \dots & 0.04390 & 8.7\tten{-4} & -317 & {\dots} & {\dots} & \dots & \dots & \dots & \dots\\
2002bf & \dots & 0.02525 & 8.6\tten{-4} & -92 & {\dots} & {\dots} & \dots & \dots & \dots & \dots\\
\bottomrule
\end{tabular}
\end{threeparttable}
\end{table*}

\subsection{Description of parameters}\label{subsec:description}
The relevant parameters for our study are the redshifts and peculiar velocities.
The heliocentric redshift (\zhel{}) is the ``observed'' redshift.\footnote{Heliocentric redshifts are not quite the observed redshift, but we assume corrections for the Earth's motion relative to the sun have been made correctly by the analysis software.
This correction is small ($\Delta z\lesssim10^{-6}$), so its effect would be negligible for current cosmological studies.}  
We convert from \zhel{} to the CMB frame redshift (\zCMB{}) using the standard formulae in Section~\ref{sec:convert2cmb} and emphasise that we do not approximate these transformations.
CMB-frame redshift refers to the redshift after we correct for only {\em our own} peculiar velocity, i.e.\ we correct for the Planck-observed CMB dipole.
The peculiar velocity (\vp{}) and corresponding peculiar redshift (\zp{}) refer to the motion of the distant galaxy that is in addition to the Hubble flow.
The final Hubble-diagram redshift that is used for cosmology (\zHD{}) is the final stage, after we have corrected \zCMB{} for the peculiar velocity of the distant galaxy.
The standard formulae are also given in Section~\ref{sec:convert2cmb}, and the derivation of the peculiar velocities themselves is described in Section~\ref{sec:pv}.
Each of these parameters have uncertainties represented by $\sigma$.
When the parameters come from the host galaxy group \citep[from][]{Peterson2021} instead of the individual host or SN, the symbol is preceded by `Group'.

\subsection{Corrections and additions to previous data}\label{subsec:corrections}
Pantheon+ carries over the same SNe from Pantheon and includes many more SNe from FSS, DES, LOSS, SOUSA and CNIa0.02, as defined in \citet{Scolnic2022samples}.
Therefore, redshift and bookkeeping mistakes are carried over from Pantheon which were in turn carried over from their original sources, mostly from older SN compilations.
After examining each of the samples listed in Table~\ref{tab:panthplus_description} we found and fixed various errors, and added improvements as follows: 
\begin{itemize}
    \item GOODS+PANS and SCP had Right Ascension (RA) and Declination (Dec) listed as zeros.  As a result, the heliocentric corrections had been made to the incorrect part of the sky.  We assign coordinates from the original datasets \citep{Riess2004, Riess2007, Suzuki2012} and recompute \zCMB{}.
    \item We provide updates to DES redshifts from their 3-year values \citep{Smith2020} to their (previously unpublished) values that will be used in DES 5-year cosmology. This includes reassigning uncertainty based on whether the redshift comes from the host or SN spectrum:  5\tten{-4} and 5\tten{-3} respectively. 
    \item All SCP and GOODS+PANS SNe had redshift uncertainties set to 1\tten{-3} regardless of redshift source, so we reassigned the uncertainties the same way as with DES.
    \item Six FSS SNe had coordinates that disagreed with both the NED entry and FSS-assigned-host by up to tens of degrees.  
    We therefore correct the SN coordinates to the NED coordinates, as listed in Table~\ref{tab:FSSpos}.
    \item One CfA4 SN had its location mistaken for a SN discovered around the same time. The record for 2008cm had the coordinates of SNF 20080514-002, but since the redshift was in agreement with the host of 2008cm, we update the SN coordinates using NED (Table~\ref{tab:FSSpos}). 
    \item We update the coordinates of CfA1 SN 1996C to those of SIMBAD because the NED coordinates are incorrect (Table~\ref{tab:FSSpos}).
    \item Where we have identified the host of a SN, host coordinates are provided separately to SN coordinates. There was previously no record of SN hosts in Pantheon or some source catalogues.
     Heliocentric corrections are performed using host coordinates where possible.
    \item Where an International Astronomical Union name (IAUC) for a SN exists, we record it alongside the internal ID (SNID). The IAUC links SNe that are common across samples that use different internal names. However, we recommend that in future, a dictionary of all names for a SN be implemented since SNe without an IAUC must be matched via position instead. 
    \item In collaboration with \citet{Peterson2021}, we include group-centre coordinates and group-average heliocentric redshifts, from which we derive group \zCMB{}, group \zHD{} and group $\vp{}$ (see Table \ref{tab:master}).
\end{itemize}

As a result of these changes, all SNe now have the same information: both the SNID and IAUC where applicable, both SN and host coordinates where applicable, redshifts in the heliocentric and CMB frame, and finally our updated peculiar velocities.

\begin{table*}[t!]
\caption{Corrections to SN positions.\label{tab:FSSpos}} 
\begin{tabular}{lS[table-format=3.6]S[table-format=3.6,retain-explicit-plus]S[table-format=3.5]S[table-format=3.5,retain-explicit-plus]}
\toprule
{} & {Prev.~RA} & {Prev.~Dec} & {Updated RA} & {Updated Dec} \\ {SNID} & {\dg{} (J2000)} & {\dg{} (J2000)} & {\dg{} (J2000)} & {\dg{} (J2000)}\\
 \midrule
1996C & 207.75158 & +49.341251 & 207.7025 & +49.31864 \\ 
2008cm & 202.30344 & +11.272403 & 109.16146 & -62.31469 \\
vespasian & 215.136083 & +53.046728 & 322.4275 & -7.69658\\
ASASSN-15ga & 208.159615 & +10.718024 & 194.86372 & +14.17105 \\
SN2016aqb & 164.193214 & -13.047481 & 170.49333 & -13.98367 \\
ATLAS16bwu & 93.620268 & -17.421361 & 18.59275 & -13.15309 \\
PS16bnz & 158.658125 & +27.190117 & 155.15375 & -2.46675 \\
PS16eqv & 53.168411 & -19.615435 & 37.93083 & -25.00162 \\
ASASSN-17aj & 101.946198 & +7.920005 & 173.29375 & -10.22177 \\
\bottomrule
\end{tabular}
\end{table*}

\section{Reviewing heliocentric redshifts}\label{sec:validhelio}
The most accurate redshift for a SN---in the absence of a galaxy group average redshift---is that of its host galaxy, so it is imperative that the correct host is assigned.
This is true at any redshift, but especially so at low redshift.
In this work, we define the low-$z$ sample as the $\sim$1000 galaxies with $z\lesssim0.12$ (median $z\sim 0.024$).
These are primarily the SNe that were not uniquely observed in the typical `high-$z$' surveys of SNLS, DES, SDSS, PS1MD and HST, however some SNe in these surveys are low-$z$ since they were also observed by the low-$z$ surveys.

In the interest of thoroughly reassessing the redshifts of Pantheon+, we used Aladin \citep{Bonnarel2000} to visually inspect Pan-STARRS images at every low-$z$ (and occasionally moderate-$z$) SN location, to assign and record hosts.
It is this requirement of visual inspection that makes our definition of low-$z$ unique.
We also used Dark Energy Camera Legacy Survey, SDSS and DES images where available, and for Dec $\lesssim-30\dg$ we used mostly Digitized Sky Survey 2 images.
For the hosts we identified, we chose the `best' redshift according to a hierarchy of criteria outlined below. 
For the three low-$z$ SNe we could not assign hosts---1996ab, 2007kg and 2008dx---we use the redshift given in the original classification reports. 
Apart from these three, there was ambiguity in only six low-$z$ hosts (see discussion below).
In addition the low-$z$ sample for which we identified redshifts, we confirmed the coordinates and host names of all SDSS and FSS SNe. 

Some redshifts in NED are supplied with a comment as to their origins, which contribute to picking the best redshift. 
The origins are either `from new data' (i.e.~the reference measured the redshift), `from reprocessed raw data', or `from uncertain origins'.
Most redshifts are from uncertain origins, usually because the sources re-record older redshifts, but also because some NED records do not report that the redshift is new.

The criteria for picking the best redshift are:
\begin{enumerate}
    \item We use SDSS Data Release 13 (DR13) redshifts when they are available, as these are usually the most up-to-date measurement, have low uncertainty, and show stability in that earlier iterations of SDSS tend to converge on the DR13 values. 
    \item Next we consider other sources that include uncertainty estimates. Among these we choose the one that first satisfies, in order of decreasing priority:
    \begin{enumerate}
    \item The most recent source that has taken new data and measured a new redshift.
    \item The most recent source that has reprocessed old data.
    \item The most recent source that has an uncertain origin. This may be original data, but may also be from a new publication that uses old redshifts because these often appear as new entries in NED. We endeavoured to avoid republished redshifts and quote the original source. 
    \end{enumerate}
    In the case of multiple redshifts satisfying any condition, we take the most precise redshift.
    \item If none of the above criteria are satisfied we consider redshifts that lack an uncertainty estimate, but are not a SN redshift. In these cases, we set the uncertainty to 9\tten{-5} (see Section \ref{sec:uncertainties}). 
    \item As the last resort we take the redshift derived from the original SN spectrum. In the cases where no redshift uncertainty is reported, we set the uncertainty to be 5\tten{-3} (Section \ref{sec:uncertainties}).
\end{enumerate}

We examined all redshifts in the low-$z$ sample except those that come from FSS.
These redshifts were {\em not} updated because FSS adopt their own hierarchical approach to selecting redshifts in the literature \citep{Foley2018}.
Importantly, FSS measure new host redshifts for SNe previously without host redshifts.
Thus, we assume this sample has the best existing redshift estimates already.

\begin{figure}[t!]
    \centering
    \includegraphics[width=\textwidth]{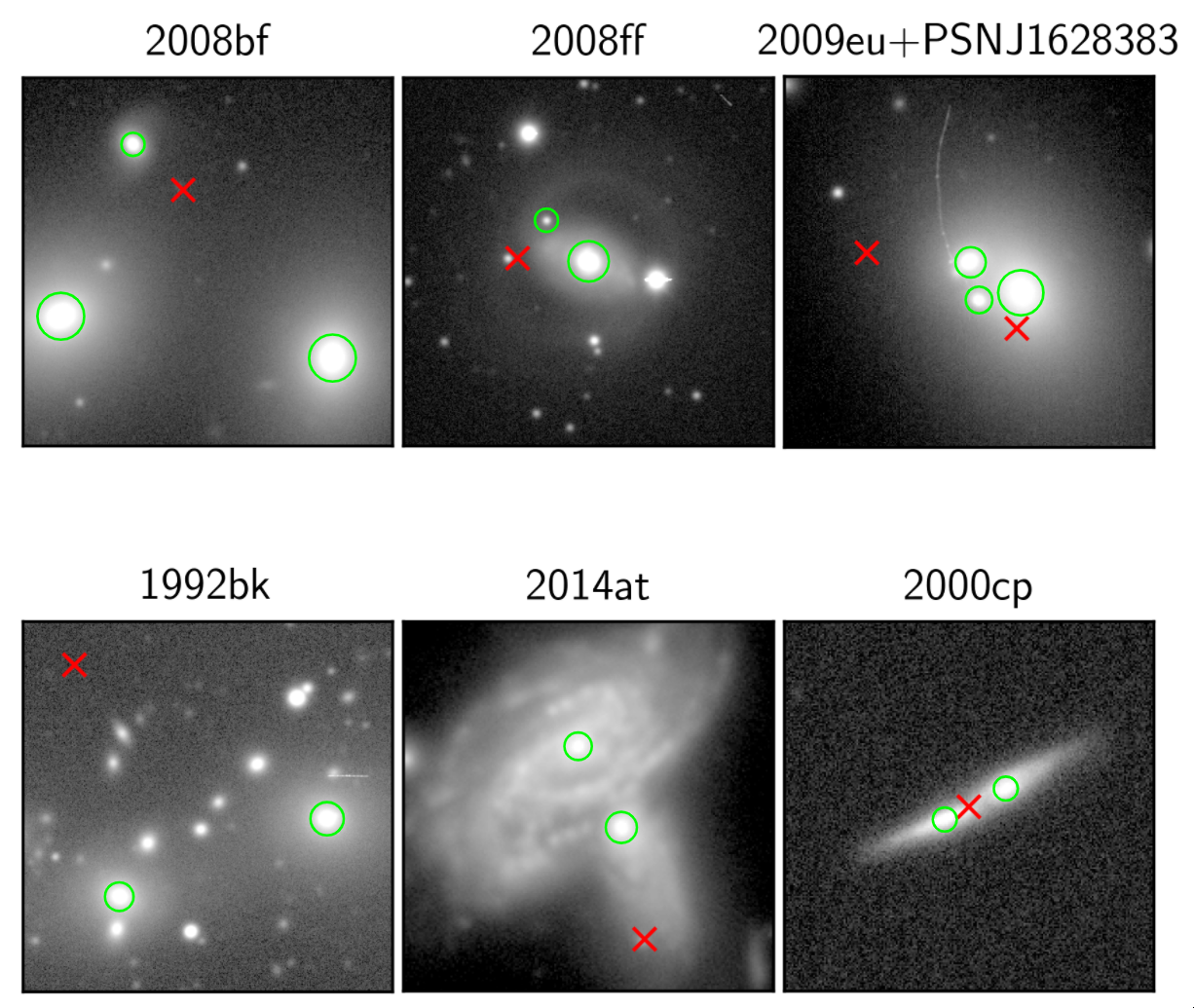}
    \caption{All instances of an ambiguous host where redshifts of multiple likely hosts have been averaged (in each case all possible hosts are at approximately the same redshift). In all images, North is up and East to the left. Red crosses indicate the location of the SN, and green circles indicate the hosts used. Other objects in the images were not selected because they were stars, or were less likely host galaxies due to their size.}
    \label{fig:host_groups}
\end{figure}

Out of 2285 unique SNe Ia, 990 heliocentric redshifts have been updated.
This is mainly for low-$z$ hosts, whose redshifts have been measured multiple times.
Some notable cases:
\begin{itemize}
    \item All HST supernovae had \zhel{} listed as \zCMB{}.  In each case, we use the heliocentric redshifts given by the original publication and recompute \zCMB{}.
    \item SN 1992bk, 2000cp, 2008bf, 2008ff, 2009eu, 2014at and PSNJ1628383 each occurred in a group of galaxies so that a unique host could not be determined. The most accurate treatment of these cases, and in fact in general, is to average the redshifts of host-galaxy-group members, as in \citet{Peterson2021}, with more members giving a more accurate redshift. 
    Figure \ref{fig:host_groups} shows images for these six cases, with SN locations and potential hosts indicated. 
    We set the group redshift uncertainty to the dispersion in group member redshifts. 
    SN 1992bk is a good example of how the `directional light radius' \citep[DLR; see][]{Gupta2016} might be used to exclude the smaller, less likely cluster galaxies from being considered as hosts. We do not use DLRs to determine the most likely host, and instead visually identify the most likely group members and take their average redshift, which gives a more accurate estimate of the Hubble-flow redshift than any single group member. 
    \item Some SNe had multiple unique SDSS redshift measurements from the same data release. The unique redshifts were averaged, and uncertainties estimated from the dispersion of redshift measurements, similar to the ambiguous host redshifts (see Table~\ref{tab:SDSSavgs}). The quoted redshift uncertainties are often less than the standard deviation of the measurements (despite the fact that standard deviations measured from small samples are typically underestimated) indicating that the redshift uncertainties may be underestimated; see discussion in Section~\ref{sec:uncertainties}. 
    \item 
    There were many cases where a SN redshift was quoted in place of a more accurate host redshift, as seen in the discrepancies between the four different redshift sources \citet[][Table 1 and Table 9]{Sako2018}, \citet{Gupta2011}, and \citet{Ostman2011}.
    We examined every image of SDSS SNe to confirm the host, SN coordinates, and attempt to update the redshift using NED.
    In ambiguous host cases, we use the host coordinates published in \citet{Sako2018}. 
    In this way, we replace 81 SN redshifts with host redshifts.
    \item \citet{Zheng2008} noted in the first year SDSS SN data release that there was a systematic offset of 3\tten{-3} between host redshifts and SN redshifts, and thus they applied the shift to SN redshifts to bring them in line with host redshifts.
    \citet{Sako2018} found a similar offset of $(2.2\pm0.4)\tten{-3}$ was present in their larger sample of SDSS SNe, however they did not apply the shift to the SN redshifts.
    We confirm that we see the same trend in the 81 SDSS SNe whose SN-$z$ we replace with host-$z$.
    Unlike \citet{Sako2018}, we \textit{do} apply this systematic redshift shift of $+2.2\tten{-3}$ to the remaining 46 SDSS SN-$z$s.\footnote{Redshifts are added together by multiplying $1+z$ factors, however here we apply the correction additively, which we believe is how the systematic offset was modelled.}
\end{itemize}

Table~\ref{tab:update_outliers} shows the largest disagreements between new and old redshifts in decreasing level of disagreement, with justifications for the update.
Since we expect updates from SN-$z$ to host-$z$ to be large, we focus on disagreements between host-$z$s.

\begin{figure}[t!]
    \centering
    \includegraphics[width=\textwidth]{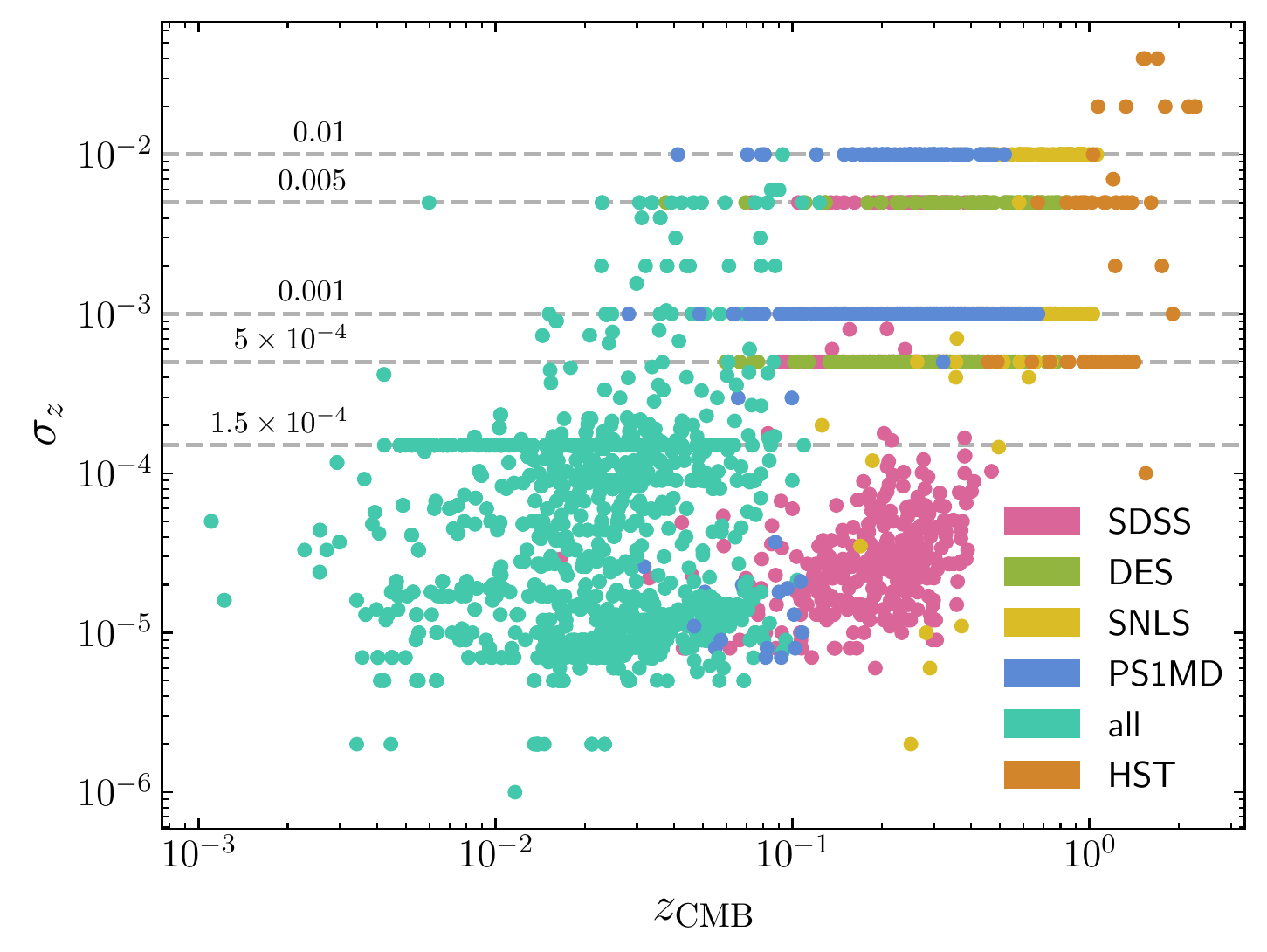}
    \caption{Uncertainty in redshift versus redshift. We observe a general upward trend with redshift as expected. Samples that use standard uncertainties have been highlighted with dashed lines. The standard uncertainty of 1.5\tten{-4} is that of the 6dF redshifts in the low-$z$ sample.}
    \label{fig:avg_err_vs_z}
\end{figure} 

\section{Estimating redshift uncertainties}\label{sec:uncertainties}

Ideally, each object would have an individual redshift uncertainty measurement based on its spectrum.
However, many surveys (generally those at higher redshift) do not provide this information and instead give overall estimates of redshift uncertainties for typical classes of object (e.g. DES, PS1MD, SNLS; see Figure \ref{fig:avg_err_vs_z}). 
There are several ways to estimate the redshift of an object: emission-line redshifting \citep[e.g.][]{Colless2001}, cross-correlation with templates \citep{Tonry1979}, least-squares minimisation \citep{Bolton2012}, or from 21 cm \Hi\ emission profiles in radio \citep[e.g.][]{Springob2005}.
Each method has its own way of estimating redshift uncertainty per spectrum, and the Pantheon+ sample has redshifts determined from each of these methods.
For instance, SDSS uses least-squares minimisation, DES/OzDES uses cross-correlation, there are many galaxies with 21 cm \Hi\ redshifts \cite[mostly from][]{Springob2005}, and many more that are unspecified.

\subsection{Typical Uncertainties}\label{subsec:typical_unc}
More problematic than reporting only class-based uncertainties is the non-reporting of uncertainties. 
Some SN data releases did not include uncertainties (e.g., CSP, CfA), in which case the only way to obtain an uncertainty estimate is to instead use an original measurement of the redshift.
Our aim is for every redshift to have an uncertainty estimate, so we must estimate uncertainties where none are reported. 
Despite identifying the best primary source according to the hierarchy above, a small number of sources did not provide uncertainties. 
With no information about the spectra themselves, we opt to set missing uncertainties to a typical value, according to whether they are host or SN redshifts.

We examine the redshift uncertainties of five surveys with redshifts in Pantheon+ or that provide their own galaxy redshift uncertainty statistics: the Two-degree Field Galaxy Redshift Survey \citep[2dFGRS;][]{Colless2001}, the WiggleZ Dark Energy Survey \citep{Drinkwater2018}, SDSS, DES/OzDES \citep{Lidman2020} and the Digital Archive of \Hi\ 21 Centimeter Line Spectra from \citet{Springob2005}.
However, we defer the discussion of SDSS-measured redshifts to Section~\ref{subsec:underest-SDSS} as SDSS redshift uncertainties are consistently smaller than other optical estimates.

While each method of redshifting naturally provides a way of estimating uncertainty, the more common way to estimate uncertainty is by comparing multiple measurements of the same object. 
Each survey found:
\begin{itemize}
    \item 2dFGRS: Over $0<z<0.3$, the RMS of multiple measurements was $\sigma_z=2.8\tten{-4}$. The authors found a slight upwards trend with redshift.
    \item WiggleZ: Over $0<z<1.3$, the standard deviation of multiple measurements ranged from $\sigma_z=1.7\tten{-4}$ for the highest quality spectra to $\sigma_z=2.7\tten{-4}$ for the lower quality but successfully redshifted spectra. The authors found no trend with redshift.
    \item OzDES: The collaboration opted to estimate uncertainties for classes of objects, assigning general SN host galaxies uncertainties of $\sigma_z=1.5\tten{-4}$, which the authors state is a lower bound.\footnote{To be conservative we have set the host galaxy redshift uncertainty for DES SNe that are in Pantheon+ to be 5\tten{-4} (M.~Smith, private communication).}
    \item \citet{Springob2005} 21 cm Archive: Compared with optical, redshifts determined from the 21 cm line offer impressive precision. After reanalysing nearly 9000 nearby ($-0.005<z<0.08$) \Hi\ galaxies, the mean and median redshift uncertainty was 1.7\tten{-5} and 1.2\tten{-5} respectively.
\end{itemize}

\begin{figure}[t!]
    \centering
    \includegraphics[width=\textwidth]{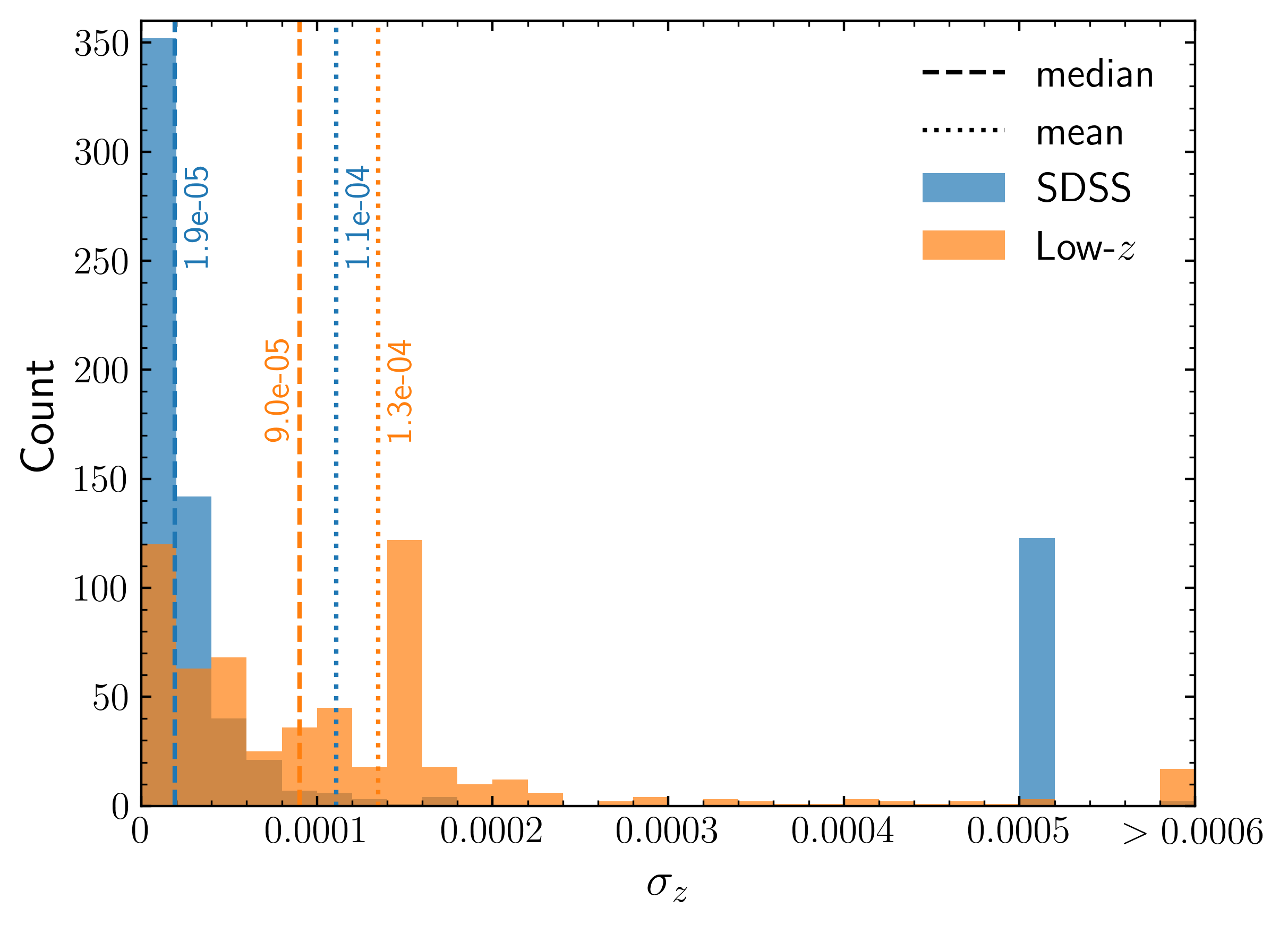}
    \caption{Comparison of SDSS DR13 redshift uncertainties and all other low-$z$ SN uncertainties. SDSS DR13 has a tight distribution peaking around 1\tten{-5}, while low-$z$ is much broader due to its heterogeneous redshift sub-samples.  
    Redshifts larger than 0.0006 have been collected into a single bin, as the maximum is 0.002 \citep{Perlmutter1999}. 
    The spike in low-$z$ at $\sigma_z=0.00015$ is due to 6dF, while the SDSS spike at $\sigma_z=0.0005$ corresponds to the uncertainty typically set for a redshift from emission/absorption lines in a SN spectrum.}
    \label{fig:avg_errs_SDSS_vs_lowz}
\end{figure}

For the Pantheon+ low-$z$ sample, which has a mix of optical and radio redshifts from the literature, we find a median redshift uncertainty of 9\tten{-5}, which is consistent with the above measurements.
We also include in this test the 30 PS1MD and SNLS SNe whose host galaxies we find third party redshifts for, extending to $z\sim0.50$.
We show in Figure \ref{fig:avg_errs_SDSS_vs_lowz} the distribution of the redshift uncertainties of this modified low-$z$ sample.
Since there remain 30 host galaxies that lack redshift uncertainties, all at low-$z$, we opt to assign the median uncertainty calculated from the rest of our low-$z$ sample (see Table~\ref{tab:missing_uncertainties}). 

Until now, we have discussed only host galaxy redshifts because these have been studied in detail. 
Estimating redshifts from SN spectra has had less attention. 
Like host galaxy redshifts, there are SN redshifts without uncertainty estimates. 
In these cases, we rely on a less concrete estimation of $\sigma_z=0.005$, based on the population of SN redshifts to date and expert opinion.
We reiterate that this is a somewhat conservative but fair estimation. 
As an example, as given by \texttt{SNID} \citep{Blondin2007b}, the mean redshift uncertainty of the 51 supernovae with Type Ia \texttt{SNID} classifications from the CfA Supernova Group website\footnote{\url{https://lweb.cfa.harvard.edu/supernova/OldRecentSN.html} and \url{https://lweb.cfa.harvard.edu/supernova/RecentSN.html}} is $\sigma_z=4.1$\tten{-3} with a standard deviation of 1.7\tten{-3}.
  
See Table~\ref{tab:missing_uncertainties} for a list of all SNe previously missing uncertainties. 

\subsection{Underestimated SDSS Uncertainties}\label{subsec:underest-SDSS}

\begin{figure}[t!]
    \centering
    \includegraphics[width=\textwidth]{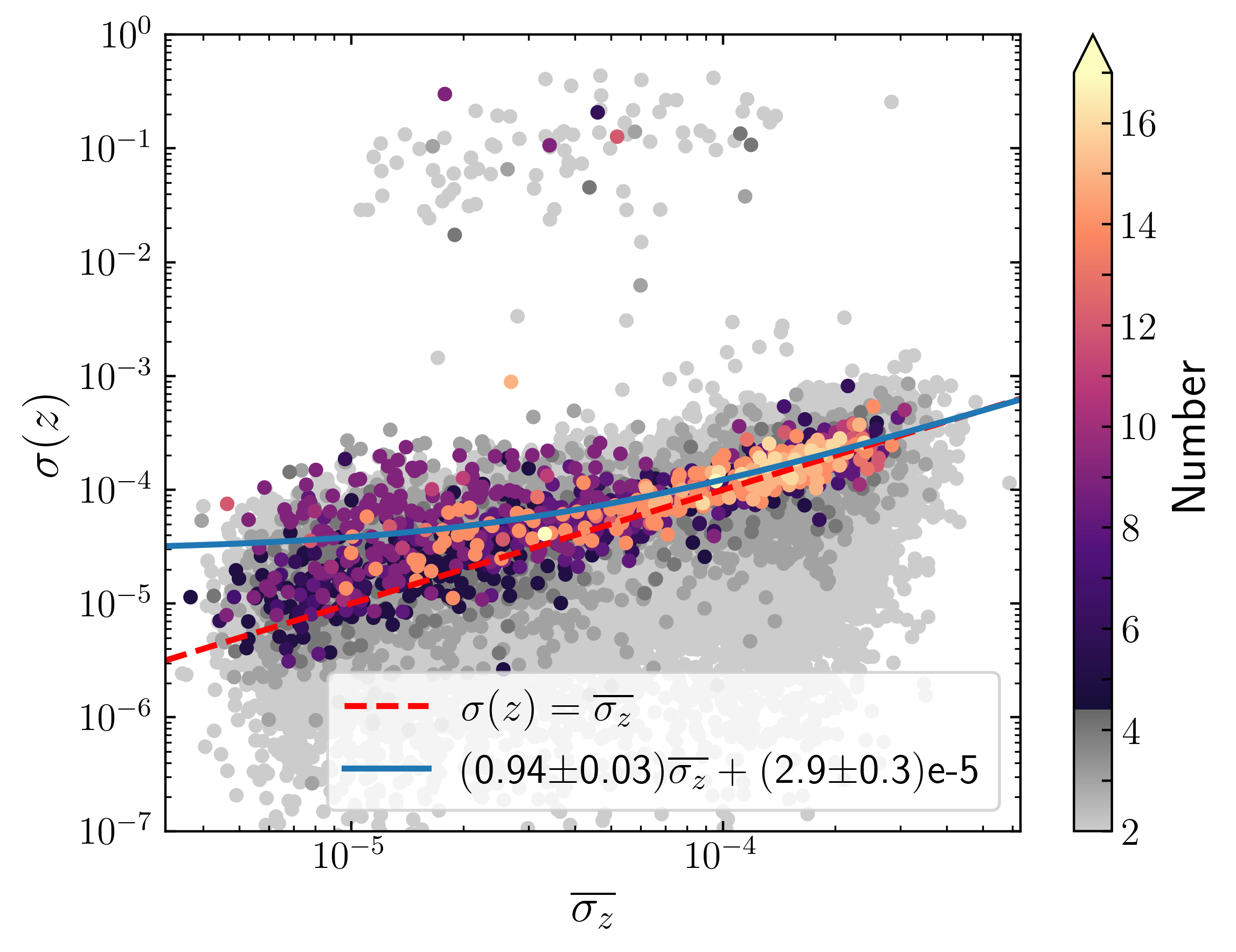}
    \caption{Standard deviation of multiple SDSS DR13 reliable redshift measurements of the same object against the average of their quoted uncertainties. We expect a slope of 1.0 (red dashed line) if the uncertainties are appropriate. The solid blue line is a linear fit to all data with five or more measurements and a dispersion of $\sigma(z)<0.01$, which shows that, consistently, $\sigma(z)>\overline{\sigma_z}$. Every point with $\sigma(z)\gtrsim0.01$ is a catastrophic redshift failure caused by at least two distinct confident redshifts.
    }
    \label{fig:SDSS_scatter}
\end{figure} 

We noted in Section \ref{sec:validhelio} that when multiple redshift measurements were available in SDSS, the dispersion in redshift values is usually higher than expected from their typical uncertainties of 1\tten{-5}.
This comparison indicates that typical SDSS redshift uncertainties are somewhat optimistic.  
We quantify this comparison by calculating both average uncertainty, $\overline{\sigma_z}$, and dispersion in $z$, $\sigma(z)$, for all multiply-measured objects in SDSS DR13.
Only objects with two or more reliable redshifts are used.

The average uncertainty versus $z$ dispersion for all multiply-measured objects is shown in Figure \ref{fig:SDSS_scatter}.
We perform a linear fit to the objects with five or more measurements weighted by number of redshift measurements.
Calculating standard deviation from few points is systematically biased low (grey points), so we exclude objects with two--four measurements.
We also exclude dispersions above 0.01 since these dispersions are caused by multiple confident but disparate redshifts.
Thus the fit was performed on 1,000 of 31,000 objects.

The uncertainty of 9\tten{-5} that we apply to sources with no provided uncertainty reflects the mean dispersion of 9.5\tten{-5} we calculate for repeated non-outlying SDSS measurements. 
Since we see a gradient of almost unity and a positive offset of roughly 3\tten{-5} between dispersion and estimated uncertainty, increasing the SDSS uncertainties by this amount would be reasonable, and we test the impact of this increase in our cosmological analysis in Section \ref{sec:impact}.\footnote{Our increased uncertainty better reflects the observed dispersion between SDSS measurements but may still underestimate the uncertainties given the typical uncertainty of optical redshifts in Section \ref{subsec:typical_unc}.}

\section{Combining redshifts multiplicatively}\label{sec:convert2cmb}
\subsection{Heliocentric corrections}
Most publicly available heliocentric corrections, including those currently in NED and Pantheon used a low-redshift approximation \citep[although the approximation in NED will be corrected; see][]{Carr2021}.
When performing the heliocentric correction, the low-redshift approximation assumes the observed redshift \zhel{} is an {\em additive} combination of \zCMB{} and the redshift due to our Sun's peculiar motion, \zSun{},
\begin{equation} \label{eq:additive} 
\zhel{}=\zCMB{+}+\zSun{}.
\end{equation}
However, the correct way to combine redshifts is to multiplicatively combine factors of $(1+z)$, so
\begin{equation} \label{eq:multiplicative} 
(1+\zhel)=(1+\zCMB{\times})(1+\zSun{}). 
\end{equation}
This gives the correct CMB-frame redshift, which is
\begin{equation} \label{eq:full_correction} 
\zCMB{\times} = \frac{1+\zhel{}}{1+\zSun{}}-1. 
\end{equation}
The difference between using the additive approximation and the correct multiplicative equation is exactly \zCMB{}\zSun{}.
Since \zSun{} is our own motion with respect to the CMB (our velocity $v_{\text{CMB}}$ in the direction of the CMB dipole), it is of order $\ten{-3}$. 
Therefore, at low-$z$, the difference between \zCMB{\times} and \zCMB{+} appears almost negligible; however, by $z\sim 1$, the error is on the order of \ten{-3}, which is an order of magnitude larger than most reported statistical uncertainties in redshifts.

We have ensured that all sub-samples in the new Pantheon+ sample consistently use the multiplicative correction.

\subsubsection{Which dipole to use?} \label{subsec:dipole}

The Pantheon sample mostly used the CMB dipole measured by the Cosmic Background Explorer (COBE) satellite \citep{Fixsen1996}, in the direction of galactic longitude and latitude $(l, b) = (264.14\dg{}\pm0.30\dg{}, 48.26\dg{}\pm0.30\dg{})$ with a velocity $\voC = 371\pm1$ \kms{}.

We update the heliocentric correction to use the dipole measured by the Planck Collaboration \citep{Planck2018}, $(l, b) = (264.021\dg{}\pm0.011\dg{}, 48.253\dg{}\pm0.005\dg{})$ with a velocity $\voP = 369.82\pm0.11$ \kms{}.
The difference in redshift between using the COBE dipole and the Planck dipole is at most $\sim\ten{-5}$, so this is a small change.

\subsubsection{Calculating \zSun{}}
The projection of the Sun's peculiar velocity along the line of sight to an object is
\begin{equation} \label{eq:dot}
    \vSun{} = \bm{v}_{\text{Sun}}\cdot\hat{\bm{n}}_{\text{obj}} = \vSun{max}\cos\alpha, \end{equation}
where $\hat{\bm{n}}_{\text{obj}}$ is the object's position vector, and $\alpha$ is the angle separating the dipole direction and the object.

Since the Sun's velocity is small (order of \ten{2} \kms{}) compared to $c$, the low-$z$ approximation $\zSun{} \approx -\vSun{}/c$ is adequate, but we use the full special relativistic calculation 
\begin{equation} \zSun{} = \sqrt{\frac{1+(-\vSun{})/c}{1-(-\vSun{})/c}}-1,\end{equation} 
because there is negligible computational advantage to the approximation.
The minus signs before \vSun{} have been left explicit to emphasise that at zero angular separation ($\alpha=0$ in Equation \ref{eq:dot}) the object should appear slightly blueshifted due the our velocity directly towards it.

\subsection{Peculiar velocity corrections}

The peculiar redshifts arising due to the peculiar velocities of the supernova host galaxies also need to be treated multiplicatively.
Equation (\ref{eq:multiplicative}) becomes
\begin{equation} \label{eq:multiplicativeHD} 
(1+\zhel{})=(1+\zHD{})(1+\zSun{})(1+\zp{}). 
\end{equation}
Here we use the `Hubble diagram redshift' \zHD{}, which is the cosmological redshift we are interested in.
This differs in our nomenclature from the CMB-frame redshift \zCMB{}, because \zCMB{} takes into account our motion but {\em not} the peculiar velocity of the source.
Combined with Equation (\ref{eq:multiplicative})  we see
\begin{equation}
\zHD{}=\frac{1+\zCMB{}}{1+\zp{}} -1.
\label{eq:correct_dist}
\end{equation}
Thus the Hubble diagram redshift requires knowledge of the SN host's peculiar redshift \zp{}, so we turn to how we derive peculiar velocities.

\section{Updating peculiar velocity modelling}\label{sec:pv}
By applying the heliocentric-to-CMB correction we have accounted for the motion of our own solar system with respect to the CMB.
However, we have not yet accounted for the peculiar velocity of the supernova's host galaxy ($\vp{}$).
Removing the redshift due to the estimated peculiar velocity of the host galaxy leaves the cosmological redshift \zHD{} (Equation \ref{eq:correct_dist}), which is the redshift needed for the Hubble diagram.

In this section, we describe the methodology for computing the peculiar velocity for each host galaxy.
Our treatment differs from the peculiar velocities used in Pantheon in the following ways:
\begin{itemize}
    \item We use the multiplicative equation for combining redshifts (Equation~\ref{eq:multiplicativeHD}). 
    \item We convert the predicted peculiar velocity field to redshift-space (Section \ref{subsec:zspace}).
    \item Outside the measured peculiar velocity field we model the residual bulk flow as a decaying function, rather than a constant external velocity (Section \ref{subsec:outside}).
    \item We flip the $\vp$ sign convention used in the Pantheon sample (and effected the same change in the SuperNova ANAlysis software \citep[\SNANA,][]{Kessler2009} as of version 11.02). Now, $\vp$ is positive when moving away from us, which is consistent with the sign of recession velocities. 
\end{itemize}
The nominal set of peculiar velocities we derive here are examined in the companion paper \citet{Peterson2021} in the context of the efficacy of different peculiar velocity samples, models and parameters of our own model on SN Hubble residuals.  

\subsection{Estimating peculiar velocities}\label{subsec:newPVs}
The most precise way to estimate peculiar velocities is to measure the density field (e.g.\ through a redshift survey) and use that to predict the expected peculiar velocity field.\footnote{Directly measuring peculiar velocities using an independent distance measurement (such as the Tully-Fisher or Fundamental Plane relation) is less precise ($\sim20$\% uncertainties) and observationally challenging. 
Due to their sparseness, direct peculiar velocity catalogues are difficult to interpolate to get reliable peculiar velocity estimates for galaxies that do not have direct distance measurements.} 
This is known as {\em velocity field reconstruction}. 
Importantly, this method does not use supernova distances, and therefore does not introduce correlations between the peculiar-velocity-corrected SN redshift and its measured distance.
The reconstruction does require an assumed cosmological model, but the cosmological dependence is weak.  
We quantify the impact of these peculiar velocity corrections on cosmological parameters in Section~\ref{sec:impact}.

\begin{figure}[t!]
\begin{center}
\includegraphics[width=\columnwidth]{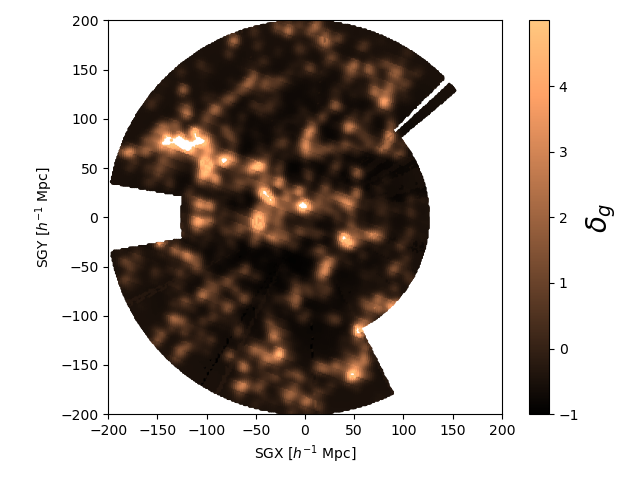}
\caption{Slice of the 2M$++$ reconstructed density field plotted in the supergalactic plane ($SGZ=0$).  White areas are regions for which data is missing, therefore the density reconstruction is uncertain.}
\label{fig:delta_star_supergal}
\end{center}
\end{figure}

\begin{figure}[t!]
\begin{center}
\includegraphics[width=\columnwidth]{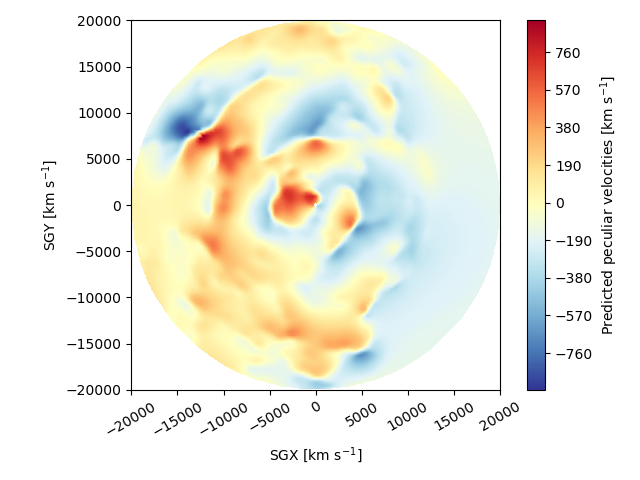}
\caption{2M$++$ velocity field plotted on a regular grid of redshift-space positions.  Note the white regions in Figure~\ref{fig:delta_star_supergal} have been interpolated over to make a complete velocity field out to 200 \hMpc{}, which means those regions of the velocity field will have higher uncertainty than other regions.}
\label{fig:Velocity_2M++}
\end{center}
\end{figure}

In the linear regime, peculiar velocity is related to the gravitational acceleration via:
\begin{equation}
    v(r) = \frac{f}{4\pi}\int \mathrm{d}^3\bm{r}^\prime \delta(\bm{r}^\prime) \frac{\bm{r}^\prime-\bm{r}}{|\bm{r}^\prime-\bm{r}|^3},
\label{v-d}
\end{equation}
where $f$ is the growth rate of the cosmic structure and $\delta(\bm{r})$ is the density contrast.
This equation has two limitations:
\begin{itemize}
    \item Galaxy surveys do not measure the total matter density, so it is assumed that the observed galaxy density ($\delta_g$) linearly traces the total density, $\delta = \delta_g/b$.  Here $b$ is the linear biasing parameter, which is different for different types of galaxies, and therefore has to be measured or marginalised over. 
    \item The region over which we have a sufficient number density of measured galaxies to do reconstruction is smaller than the region for which we need to estimate peculiar velocities.  This has been addressed by estimating the ``external velocity'', \Vext{}, which arises due to structures outside the survey volume, and estimating how that would theoretically decay with distance (Section \ref{subsec:outside}).
\end{itemize}
We use the velocity field reconstruction created by \cite{Carrick2015}, which uses data from the 2M$++$ compilation from \citet{Lavaux_and_Hudson2011}.  
2M$++$ includes data from 2MRS \citep{Huchra2005}, 6dFGS \citep{Jones2009}, and SDSS \citep{Abazajian2009} and extends to a radius of $r_{\text{max}}=200$ \hMpc{}.
One slice ($SGZ=0$) of the reconstructed density field is shown in Figure \ref{fig:delta_star_supergal} in the supergalactic plane ($SGX-SGY$).
Figure \ref{fig:Velocity_2M++} shows the 2M$++$ velocity field in redshift-space on a regular spatial grid.

A key model parameter to evaluate is $\beta\equiv f/b$.
The rate of growth is often parameterised by $f=\om^\gamma$, where $\gamma=0.55$ in the standard cosmological model, $\Lambda$CDM.
Importantly, however, $\beta$ is determined empirically rather than computed from the $\Lambda$CDM model.

Both $\beta$ and \Vext{} are derived from a combination of density field reconstruction and observation.
The reconstruction process delivers a normalised peculiar velocity field, $v_{\rm p,recon.}(\bm{r})$, which gives the directions and {\em relative} magnitudes of the peculiar velocities as a function of position.
Predictions from this peculiar velocity field are compared with galaxies that have peculiar velocities derived from distance measures such as the Tully-Fisher relation or Fundamental Plane relation. 
The calibrated peculiar velocity field is
\begin{equation}\label{eq:vp_beta_Vext}
v_{\rm p}(\bm{r}) = \beta v_{\rm p,recon.}(\bm{r}) + V_{\text{ext}}(\bm{r}).
\end{equation}
The parameter $\beta$ thus acts as a scaling of the normalised velocity field (subject to the sample of observed \vp{}), and \Vext{} is the residual mean velocity.

\cite{Carrick2015} measured $\beta=0.431\pm0.021$ and an external velocity of |\Vext{}$| = 159\pm23$ \kms\ in the direction of $(l, b) = (304\dg\pm11\dg, 6\dg\pm13\dg)$.  
While we use the velocity field measured by \citet{Carrick2015}, we use an updated value of $\beta=0.314^{+0.031}_{-0.047}$ derived in \citet{Said2020}, which gives a better fit when comparing SDSS Fundamental Plane peculiar velocities to the predicted peculiar velocity field.  
We confirm that this lower $\beta$ value results in a lower scatter in the supernova Hubble diagram, see \citet{Peterson2021}.  

\cite{Carrick2015} estimated the peculiar velocity uncertainty to be 250~\kms\ for the galaxies (the particle velocity field) and 150~\kms\ for galaxy groups (the haloes).  
In other words, the uncertainty on an individual galaxy's peculiar velocity is higher than the uncertainty on peculiar velocity of the group in which it resides.  
We note, however, that some regions of the reconstruction are less certain than others because of incomplete sampling.  
Unfortunately, sampling is not accounted for in current models, so we adopt the value of $\sigma_{v_{\rm p}}=250$~\kms\ and leave a more precise estimate of the peculiar velocity uncertainty for future work. 

\subsection{Real vs redshift-space}\label{subsec:zspace}

\cite{Carrick2015} provide the velocity field in ``real-space'', so the position and distance of a galaxy can be used to draw the peculiar velocity directly from the modelled velocity field.  
However, in supernova cosmology, the distance measured via the distance modulus should be independent of the redshift.
This distance should {\em not} be used to predict the peculiar velocity to correct the redshift.
Converting the observed redshift to distance (by assuming a cosmology) to estimate the peculiar velocity is valid, but less precise than using the redshift.
We therefore convert the reconstructed velocity field of \citet{Carrick2015} to redshift-space.  
While we assume a cosmology for this conversion, any reasonable choice has a negligible effect on the velocity field.
Thus we query the peculiar velocity model using the coordinates of each host galaxy or SN (RA, Dec, \zCMB{}), and Equation \ref{eq:vp_beta_Vext}.

The conversion to redshift-space takes two steps.
First, for each real-space grid point $i$ we convert the real-space position, $\bm{r}_i$, into redshift position $\bm{z}_i$ using the predicted peculiar velocity at that grid point, $\bm{v}_{\text{p},i}$.
Second, we use inverse distance weighting to interpolate and adjust the irregularly-spaced grid in redshift-space to a regular grid.
This process is described in more detail in Appendix \ref{sec:vpecgrids}.

\begin{figure}[t!]
    \centering
    \includegraphics[width=\columnwidth]{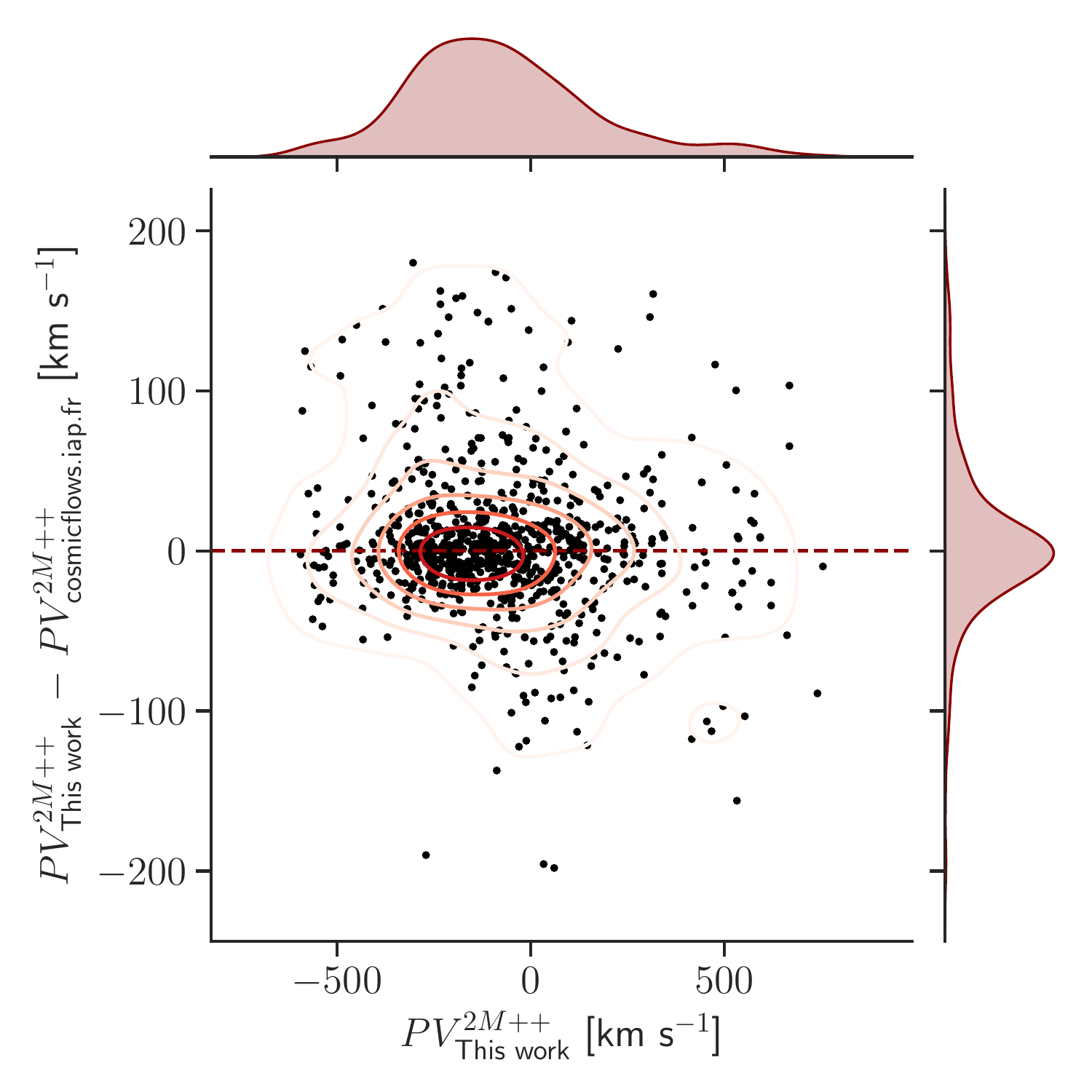}
    \caption{The predicted peculiar velocity for a sample of 851 host galaxies using two different approaches. Each \vp{} has an uncertainty of 250 \kms. We use the 2M++ velocity field converted from real-space to redshift-space whereas the method associated with \citet{Carrick2015} (previously used for Pantheon), integrates over real-space along the line of sight for each host galaxy. There are only negligible systematic differences between the results of these techniques, and all scatter is within 1$\sigma$.}
    \label{fig:pv_method_1_vs_2_hist}
\end{figure}

An alternative method is to integrate along the line of sight over real-space.
This technique is used by the online tool\footnote{\url{https://cosmicflows.iap.fr}} associated with \citet{Carrick2015}, which was previously used to estimate the Pantheon peculiar velocities. 
Both methods agree very well, within the uncertainty, as seen in  Figure \ref{fig:pv_method_1_vs_2_hist}.  
The mean difference is only 5 km s$^{-1}$ (50 times smaller than the individual uncertainty). 

\subsection{Velocities beyond $r_{\rm max}$}\label{subsec:outside}
\begin{figure}[t!]
    \centering
    \includegraphics[width=\textwidth]{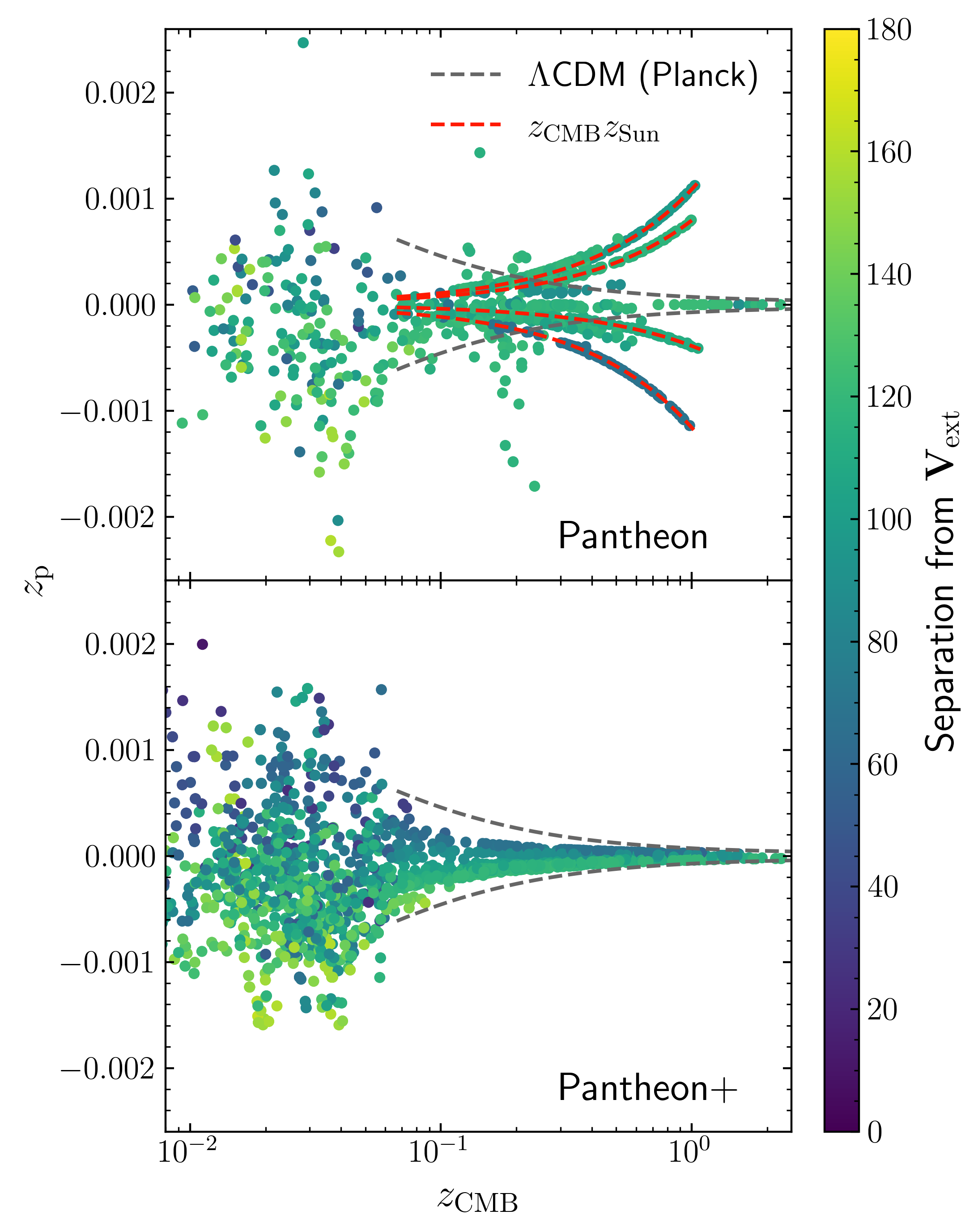}
    \caption{\textbf{Top}: Pantheon (original) peculiar velocities converted to redshift. The expected decay of peculiar velocity amplitude according to $\Lambda$CDM is over-plotted (grey dashed), outside the limit of the reconstruction at 200 \hMpc{} ($z\approx0.067$). The spurious increase in peculiar velocities as redshift increases is driven by the erroneous use of the low-redshift approximation (Equation \ref{eq:additive}). The error term is plotted (red dashed) only for the four SNLS fields. \textbf{Bottom}: Pantheon+ (this work) peculiar velocities, converted to redshift, which are now well-behaved beyond low-redshift. }
    \label{fig:pvoldnew}
\end{figure}

It is difficult to properly account for velocities outside $r_{\rm max}$ because we do not have an adequate measurement of the density field to predict individual velocities precisely.  
However, we expect velocities to continue to behave largely according to the bulk flow trend beyond $r_{\rm max}$ as a consequence of $\Lambda$CDM large scale structure.
In standard $\Lambda$CDM a theoretical bulk flow magnitude of $\sim20$ \kms\ is expected even for a sphere with radius $z\sim1$ (grey dashed lines in Figure \ref{fig:pvoldnew}).
Accordingly, peculiar velocities of galaxies outside $r_{\rm max}$ should not be set to zero.

To ensure a smooth transition across $r_{\rm max}$ we have chosen to model the bulk flow as a decaying function consistent with $\Lambda$CDM expectations, and in the direction of the bulk flow of the 200 \hMpc{} sphere.
While there is a $\Lambda$CDM model dependence, the impact of this high-$z$ correction on cosmological inferences is small both because the corrections are small (at most $\sim 5\tten{-4}$ when in the direction of the bulk flow), and because at high-$z$ these peculiar redshifts represent a small fraction of the total redshift.

We note that there is a slight difference between \Vext{} and the {\em bulk flow} of the 200 \hMpc{} sphere. 
The bulk flow of the sphere is the average of the internal velocities (which is small but non-zero), plus the external velocity. 
We calculate the bulk flow at the 2M$++$ maximum radius of 200 \hMpc{} to be $182\pm23$ \kms\ in the direction of $(l, b) = (302\dg\pm10\dg, 2\dg\pm9\dg)$. 
At this large radius, the bulk flow is dominated by the external bulk flow ($V_{\text{ext}} \approx 170$ km s$^{-1}$: \citealt{Said2020,Boruah2020}) plus a small contribution from the mean internal velocities.

Contrary to common expectations, the bulk flow should {\em not} necessarily converge on the direction of the CMB dipole. 
The observed CMB dipole is particular to our own motion, and is removed with the correction from the heliocentric to the CMB frame.  
Dramatic changes to the the magnitude and direction of bulk flow direction become unlikely as we average over spheres that approach the scale of homogeneity.
Therefore, we fix the direction of the decaying bulk flow in the direction of the 200 \hMpc{} sphere's bulk flow. 

The Pantheon sample peculiar velocities outside the velocity reconstruction suffered three main issues which can be seen in Figure \ref{fig:pvoldnew}.
The most apparent issue is the increasing \vp{} with redshift.
We show that this artefact is caused by the low-$z$ approximation of the heliocentric correction by plotting the error term (i.e. the difference between Equations \ref{eq:additive} and \ref{eq:multiplicative}) for the four SNLS fields, as these stretch to high-$z$. 
This particular error also occurs for PS1MD and SDSS but is less visible since these surveys do not extend as far in redshift.
Also visible in Figure \ref{fig:pvoldnew} are the HST SNe with $\vp{}=0$, and the somewhat random scatter around $z\approx0.15$ which may have been due to assigning 2M++ peculiar velocities outside the 2M++ sphere.
We show in the bottom panel of Figure \ref{fig:pvoldnew} that these issues are now resolved.

\section{Impact on cosmological parameters}\label{sec:impact}

\begin{table*}[ht!]
\caption{Impact of different corrections, redshift samples, systematics, and uncertainties on cosmological parameters. We define the change in cosmological parameters to be the variation minus the nominal value (variation 5). $N_{\rm SN}$ refers to the number of supernovae in each sample, which differ between variations depending on which supernovae pass or fail quality cuts as their redshifts change.  The uncertainties ($\sigma_{H_0}$ and $\sigma_w$) show only the uncertainty due to the supernova sample size and distance moduli uncertainties, not the expected precision of the measurement.\label{tab:results}}
\begin{tabular}{cp{4cm}cccS[table-format=-1.2]S[table-format=1.2]S[table-format=-1.3]S[table-format=1.3]}
\toprule
&  & \multicolumn{3}{c}{$N_{\text{SN}}$} &  &  &  &  \\ \cline{3-5} Variation & Variation Description & Total & $0.0233<\zHD{}<0.15$ & $\zHD{}>0.01$ & {$\Delta H_0$} & {$\sigma_{H_0}$} & {$\Delta w$} & {$\sigma_w$}\vspace{5pt} \\
\midrule
0 & None & 1764 & 504 & 1653 & -0.12 & 0.20 & 0.003 & 0.045\\
1 & New \zhel{} & 1764 & 500 & 1653 & -0.03 & 0.20 & 0.009 & 0.046\\
2 & New \zhel{}, \zCMB{\times} & 1763 & 500 & 1652 & -0.02 & 0.20 & 0.004 & 0.046\\
3 & \zCMB{\times}, new \vp{} & 1763 & 512 & 1653 & -0.06 & 0.19 & -0.008 & 0.045\\
4 & New \zhel{}, new \vp{} & 1764 & 512 & 1654 & -0.00 & 0.19 & 0.005 & 0.043\\
\textbf{5} & \textbf{Final (all corrections)} & \textbf{1763} & \textbf{512} & \textbf{1653} & \multicolumn{1}{r}{\textbf{0}} & \textbf{0.19} & \multicolumn{1}{r}{\textbf{0}} & \textbf{0.044}\\
\midrule
6 & Final, only host-$z$ & 1576 & 495 & 1466 & 0.05 & 0.20 & -0.022 & 0.048\\
7 & Final, group-$z$, \vp{} & 1764 & 514 & 1650 & 0.05 & 0.18 & -0.011 & 0.042\\
8 & Final, all $z-\sigma_{z}$ & 1764 & 507 & 1650 & -0.18 & 0.19 & 0.011 & 0.045\\
9 & Final, all $z+\sigma_{z}$ & 1766 & 513 & 1660 & 0.20 & 0.19 & -0.015 & 0.045\\
10 & Final, all $z-4\tten{-5}$ & 1765 & 512 & 1655 & -0.05 & 0.19 & 0.013 & 0.044\\
11 & Final, all $z+4\tten{-5}$ & 1762 & 513 & 1652 & 0.06 & 0.19 & -0.012 & 0.044\\
12 & Final, all $\sigma_{z}\times$ 3 & 1708 & 512 & 1598 & 0.06 & 0.19 & -0.014 & 0.045\\
13 & Final, SDSS $\sigma_{z} + 3\tten{-5}$ & 1763 & 512 & 1653 & 0.00 & 0.19 & 0.000 & 0.044\\
\bottomrule
\end{tabular}
\end{table*}

To test the impact of these redshift updates on cosmological parameters we fit for $H_0$, and (separately) the dark energy equation of state $w$, in a flat-$w$CDM model for various different combinations of updates as listed in Table~\ref{tab:results} and described below.
We only report the {\em changes} in these parameters, relative to the nominal `Final' set that includes all of the updates (updated \zhel{}, exact formula for combining redshifts, and new peculiar velocities). 
The full Pantheon+ cosmology analysis is reported in \citet{Brout2022cosmo}.  

In addition to combinations of redshift updates, we consider other redshift/sample variations for a total of 14 variations.  
Each variation is numbered, as listed in Table~\ref{tab:results}, and the same numbering is also included in each figure for easy reference.  
The variations we consider are:
\begin{description}
\item[(0) No corrections] This is the original data without any redshift corrections.
\item[(1--4) Partial corrections] These variations are the permutations of (a) updating \zhel{}; (b) combining redshifts multiplicatively, \zCMB{\times} (as opposed to using the low-$z$ additive approximation); (c) using our new peculiar velocities, \vp{}. 
\item[(5) All corrections] The nominal Final data includes all redshift updates.  
\item[(6--7) Redshift source] We consider first the subset of supernovae that have host-galaxy redshifts (1576 of the 1763 redshifts). 
Second, for the entire sample we replace host-galaxy redshifts with the redshift of the host galaxy's group when available (186 of 1763 redshifts).   
\item[(8--11) Systematic offsets] We consider two different forms of systematic redshift offsets: shifting the redshift by $\pm \sigma_z$ (variations 8, 9), and by $\pm 4\tten{-5}$  \citep[as suggested by][]{Calcino2017, Brout2019} (variations 10, 11).
\item[(12--13) Uncertainty changes] The last test is how the size of redshift uncertainties affects cosmological parameters \citep[as suggested by][]{Steinhardt2020}, so we scale all uncertainties up by a factor of 3 (variation 12), and only SDSS-measured redshift uncertainties by $+3\tten{-5}$ (variation 13; see Section \ref{sec:uncertainties}).  
\end{description}
In addition, in Section \ref{subsec:SNonly} we discuss the sub-sample of SNe that lack host-$z$ (187 SNe; the complement to variation 6), but we do not list this as a numbered variation due to the small number of SNe.

\begin{figure*}[ht!]
    \centering
    \includegraphics[width=\textwidth]{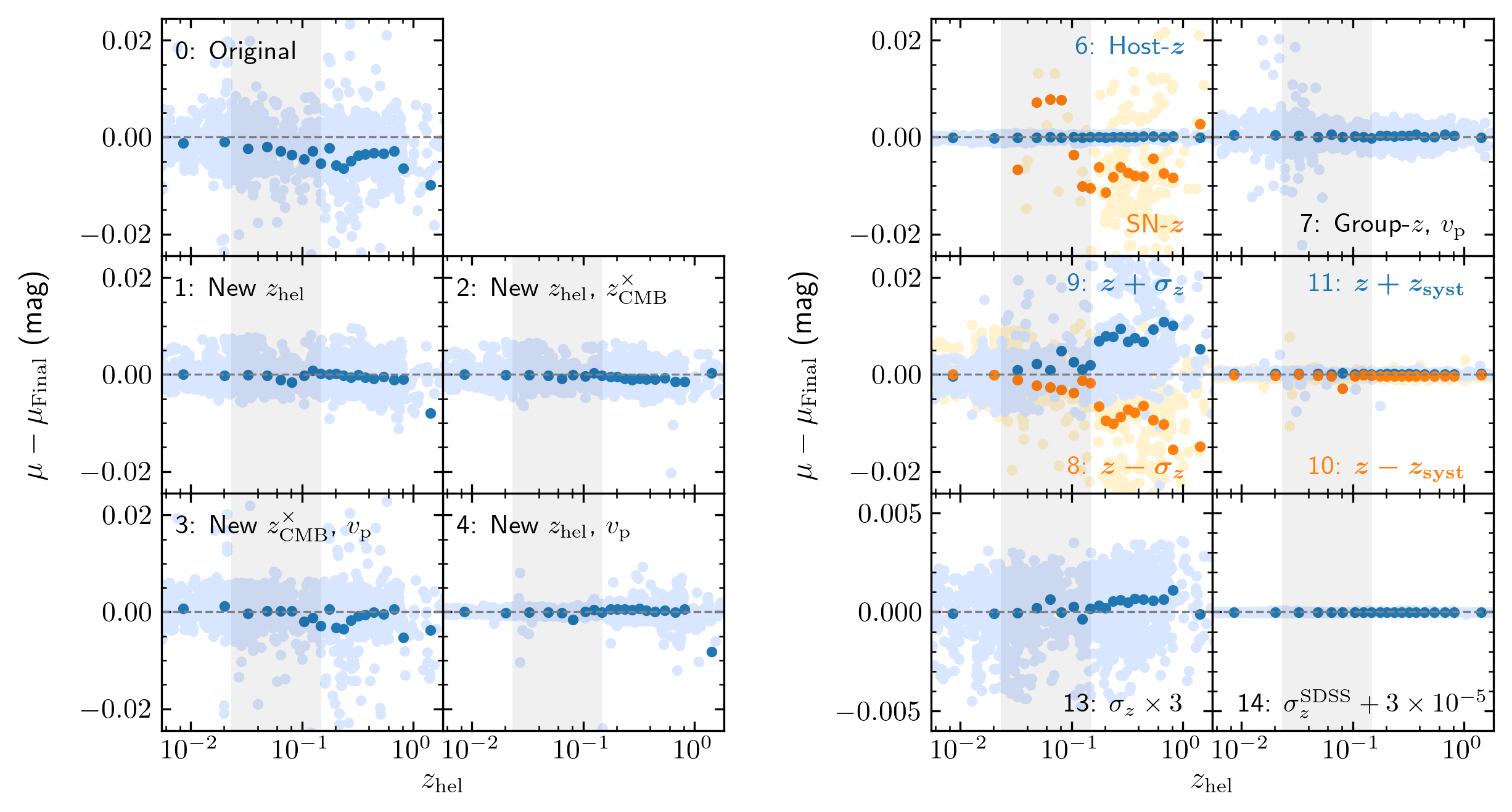}
    \caption{Difference in distance moduli for selected variations compared to the Final values (variation 5), $\mu_{\text{Final}}$.  Faint points are individual SNe, while dark points are binned in redshift.  The grey shaded region represents the redshift range used to fit $H_0$. All deviations in the binned differences are within 1$\sigma$, where the uncertainty comes only from distance modulus uncertainty (and not absolute magnitude calibration). The largest changes come from the updates to \zhel{} and next largest from the updates to \vp{}.}
    \label{fig:mudiff_all}
\end{figure*}

\subsection{Dependence of distance modulus on redshift change}\label{subsec:dmudz}
We analyse each variation independently, meaning that for each:
each SN light curve is fit using the SALT2 model \cite{Guy2010} as derived in \citet{Brout2022cals} using \SNANA\ \citep{Kessler2009}, biases are corrected following the BEAMS with Bias Correction (\texttt{BBC}) framework established in \cite{Kessler2017}, and distance moduli $\mu$ are determined, all within the \Pippin\ framework \citep{Hinton_and_Brout2020}.
The details of this process are given in \citet{Brout2022cosmo}. 
The resulting distance moduli changes can be see in Figure \ref{fig:mudiff_all}.

The total number of supernovae in each variation changes slightly due to various reasons. 
When we restrict the redshift range for fitting $H_0$ and $w$, some borderline-redshift supernovae are shifted in and out of the sample due to redshift changes.
However, shifting redshifts also affects the light curve fit parameters.
The quality cuts we apply are to these fit parameters (among others) and also their errors.
A redshift shift can therefore also shift a supernova in or out of the sample.
For example, a light curve parameter may pass the cut with the shifted redshift, or the fit may be poorer at the shifted redshift.

We expect $\mu$ to change when a supernova redshift changes because the duration of a light curve (stretch) is affected by time-dilation and its colour is affected by K-corrections, both of which are dependent on the measured heliocentric redshift (but not the peculiar velocity correction); the theoretical basis for these variations is explained in \citet{Huterer2004}. 
However, the distance modulus of a supernova may also change between sample variations without changing the redshift.
The procedure for deriving distance moduli for a supernova sample \citep[e.g.~using \texttt{BBC;}][]{Kessler2017} determines the peak magnitude ($m_B$) from the light curve, the global parameters $\alpha$ and $\beta$ that adjust the stretch ($x_1$) and colour ($c$) of the supernovae, and a correction term for selection biases ($\delta\mu_{\text{bias}}$), according to a modified version of the Tripp relation \citep{Tripp1998}, following \citet{Kessler2019},\footnote{In Equation \ref{eq:tripp} the parameters $m_B$, $\delta_{\text{bias}}$, $x_1$, and $c$ are individual to each supernova, while the parameters $M$, $\alpha$, and $\beta$ are common to the whole population.}
\begin{equation} 
\mu = m_B-M+\alpha x_1 - \beta c -\delta\mu_{\text{bias}}.\label{eq:tripp}
\end{equation} 
Therefore the distance modulus may change even if the redshift is not altered since the calibration of $\alpha$, $\beta$ and absolute magnitude ($M$) depend on the sample as a whole.
This explains the slight changes in distance moduli for variation 6 (the host-$z$ sample), in which we do not alter any redshifts. 

\begin{figure}[t!]
    \centering
    \includegraphics[width=\textwidth]{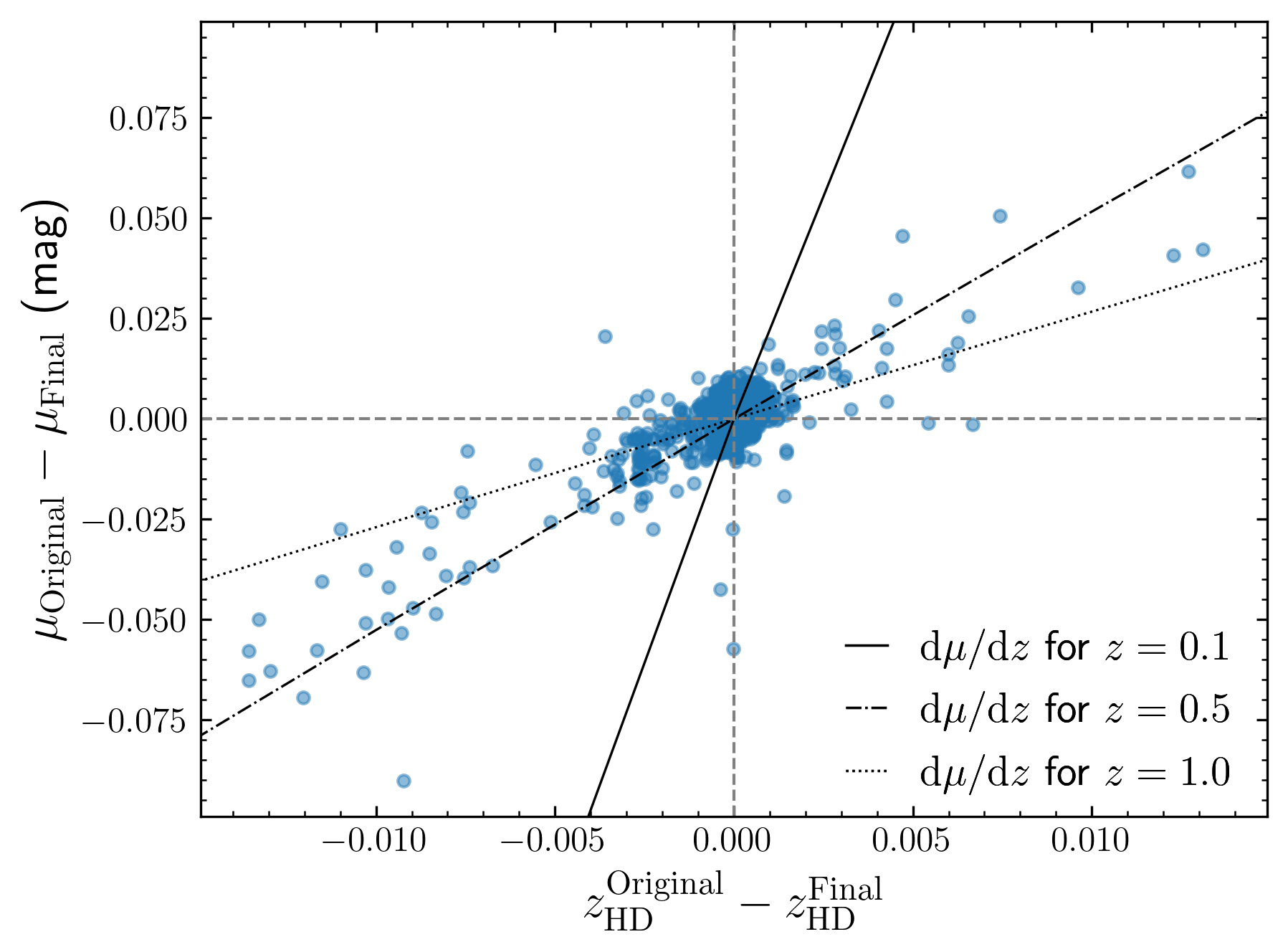}
    \caption{Change in distance modulus versus change in redshift (circles), and $\mu(z)$ derivatives d$\mu$/d$z$ for given redshifts (black lines). The largest changes occurred due to amending \zhel{}. The black lines represent the purely theoretical $\Delta\mu$ that would occur on the Hubble diagram given the $\Delta z$ on the horizontal axis (in other words, the slope of the Hubble diagram at those redshifts). The positive trend demonstrates that a change in redshift is partially cancelled the corresponding change in $\mu$ (see discussion in Section \ref{subsec:dmudz}). This trend is seen in each variation.
    }
    \label{fig:dmudz}
\end{figure}

The dependence of the change in a supernova's $\mu$ between variations on the change in $z$ is demonstrated in Figure~\ref{fig:dmudz}, which shows that the two are strongly correlated \citep[see also][]{Huterer2004}.
This correlation results in a cancellation that reduces the impact of redshift uncertainties, particularly at mid-range redshifts (around $z\sim0.5$). 
This can be seen in the solid lines in Figure \ref{fig:dmudz}, which show the slope (d$\mu/$d$z$) of the Hubble diagram at different redshifts.
At mid-range redshifts the slope is the same as the degeneracy direction between redshift change and distance modulus change. 
Thus uncertainties in redshift essentially cancel out at these redshifts, as they cause points to be shifted {\em along} the magnitude-redshift relation instead of deviating from it. 
This may be particularly helpful for supernova cosmology using photometric redshifts whose uncertainties are larger than spectroscopic redshifts \citep{Chen2022}. 

For a simplified assessment of how redshift changes impact cosmology, it is natural to simply use the published $\mu$ values, and shift the redshifts \citep[as was done in][]{Davis2019,Steinhardt2020}. 
For the low-$z$ sample, we show using the large, transparent symbols in Figure \ref{fig:H0w} that this gives a reasonable approximation to the full analysis that uses recalculated $\mu$ values (smaller symbols).  
The main difference of not recalculating $\mu$ is less cancellation of the effect of systematic redshift changes.  
At intermediate redshifts one might expect that neglecting the cancellation in d$\mu/$d$z$ may overestimate the deviation due to redshift shifts.  
Interestingly, we find that when fitting for $w$ without re-deriving $\mu$ the change in $w$ often becomes slightly larger (e.g.\ variations 0--2) but in some cases (e.g.\ variations 8--9, shifting the redshifts by 1$\sigma_z$) we find the shift in $w$ goes in the opposite direction (likely due to $\sigma_z$ increasing with $z$).  
Overall we find that doing approximate cosmology fits by changing $z$ without changing $\mu$ gives reasonable results, but for precision cosmology one should refit $\mu$ whenever $\zhel{}$ changes. 

\subsection{Fitting $H_0$}\label{subsec:H0}
We first fit $H_0$ using the method in \citet{Riess2016}, that compares the distance modulus data to a low-redshift approximation of the recession velocity.\footnote{$v(z,q_0,j_0)\approx \frac{cz}{1+z}\left[1+\frac{1}{2}(1-q_0)z-\frac{1}{6}(1-q_0-3q_0^2+j_0)z^2\right]$, which uses the canonical $\Lambda$CDM value of the deceleration parameter $q_0=-0.55$ and jerk $j_0=1$.}
The only free parameter in this fit is $H_0$.

For this fit we focus on only the low-redshift regime of $0.0233<\zHD{}<0.15$, which is the standard range used by previous supernova cosmology analyses such as \citet{Riess2016,Riess2018}.
In the Final dataset this redshift range contains 512 SNe of which only 17 lack host galaxy redshifts.

\begin{figure*}[ht!]
    \centering
    \includegraphics[width=0.65\textwidth]{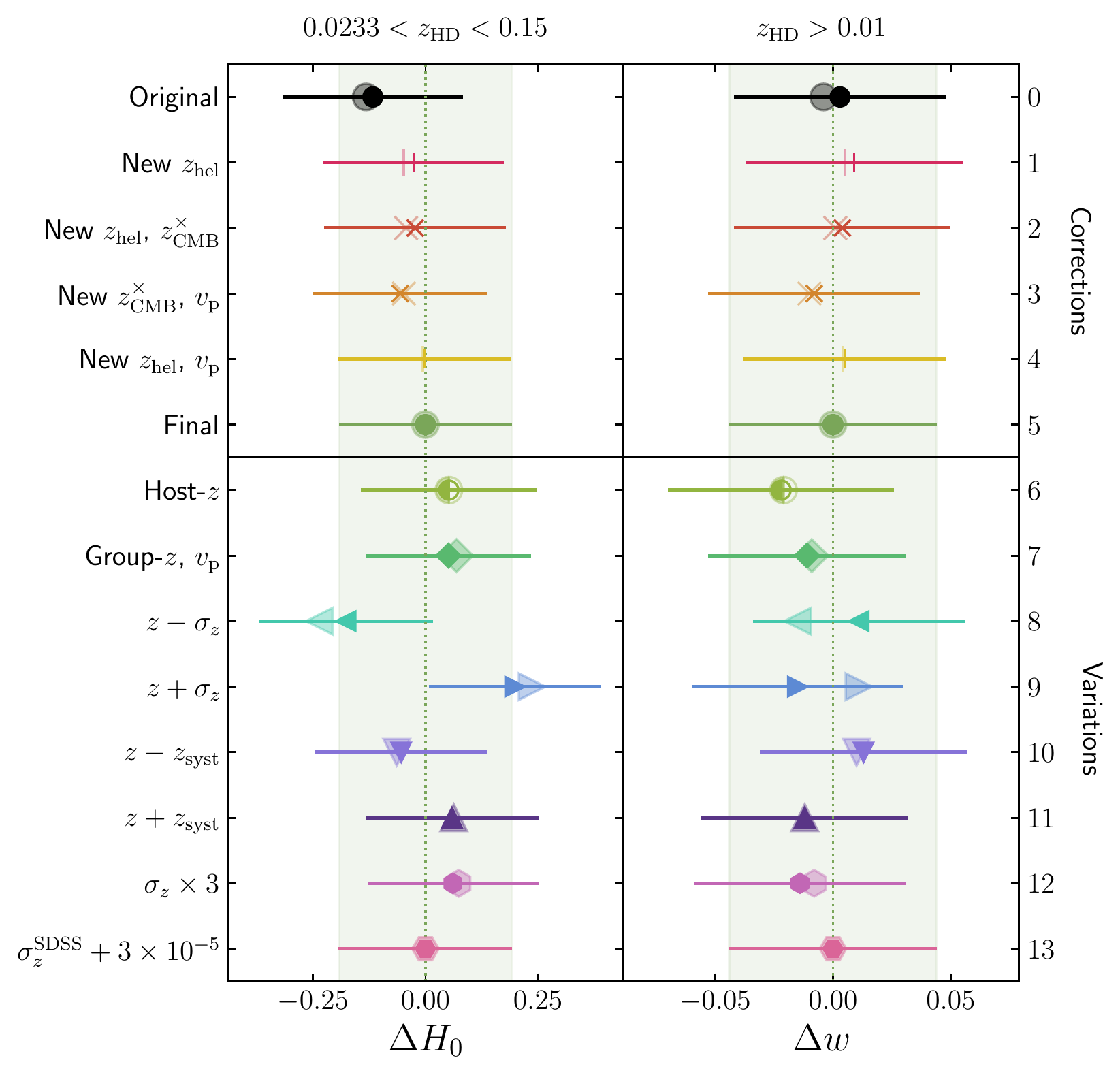}
    \caption{The impact on inferences of $H_0$ and $w$ due to changes in the redshifts. The regular symbols are the nominal results while the faint, larger symbols come from not recalculating $\mu$ after changing redshift so that the partial cancellations seen in Figure~\ref{fig:dmudz} are not present.
    The descriptions on the left hand side and variation numbers on the right hand side both correspond to the descriptions in Table \ref{tab:results}. The $H_0$ fit uses the redshift range $0.0233<\zHD{}<0.15$, while the $w$ fit uses the range $\zHD{}>0.01$, where the maximum redshift of the sample is 2.26.
    The uncertainties in both $H_0$ and $w$ only represent the statistical uncertainty from the SN magnitude uncertainties---they do not contain any uncertainty from distance ladder calibration nor intrinsic dispersion in SN magnitudes.  All variations are small, showing that the cosmology results are robust to small changes in redshift and there is no indication of systematic redshift errors biasing previous results.}
    \label{fig:H0w}
\end{figure*}

When calculating the uncertainties we only consider the statistical uncertainties in the distance moduli of the supernovae, not the uncertainties in the absolute magnitude calibration.
Thus the uncertainties in $H_0$ in Figure \ref{fig:H0w} reflect the size of shifts due only to redshift/sample variations and not the size of uncertainties in the $H_0$ measurements (for example the current uncertainty on $H_0$ from SN cosmology is about 5 times larger). 

The results are shown in Table~\ref{tab:results} and Figure~\ref{fig:H0w}, where we see that the redshift improvements we have made have a small impact on the value of $H_0$ relative to the uncertainty from the SH0ES $H_0$ measurement with uncertainty of 1.0 \kmsMpc\ \citep{Riess2022}. 
We define the difference in cosmological parameters for each variation to be the variation minus the Final value, i.e.~$\Delta H_0 = H_0^n - H_0^{\text{Final}}$ for variation $n$.  
Applying all updates to the original redshifts results in $\Delta H_0 = -0.12$ \kmsMpc{}.
The largest redshift variations, i.e. shifting all redshifts by their $1\sigma$ uncertainties (variations 8 and 9) result in $|\Delta H_0|\leq0.2$ \kmsMpc.

\subsection{Fitting $w$CDM}
We also evaluate the impact of the redshift updates on the best fit parameters in the flat-$w$CDM model.
This model has two free parameters: the matter density $\om$, and the equation of state of dark energy $w$, but we implement a Planck-like prior on the matter density of $\om=0.311\pm0.010$ so we can isolate the impact on $w$.
We marginalise over absolute magnitude and $H_0$, and we fit over the redshift range $\zHD{}>0.01$.
To estimate the changes in $w$ we use the fast minimisation routine \texttt{wFit} in SNANA \citep{Kessler2009} which outputs marginalised cosmology parameters $w$ and $\Omega_M$; the complete fit with a thorough covariance analysis can be found in \citet{Brout2022cosmo}. 

The results are shown in Table~\ref{tab:results} and  Figure~\ref{fig:H0w}; changes in $w$ are smaller than the statistical uncertainty of $\lesssim$0.03 given in the full Pantheon+ $w$ measurement from \citet{Brout2022cosmo} and the uncertainty of 0.04 from Pantheon \citep{Scolnic2018}.  
Applying all updates to the original redshifts results in $\Delta w = 0.003$.
The largest redshift variations, i.e.~shifting all redshifts by their $1\sigma$ uncertainties (variations 8 and 9) result in $|\Delta w|\leq0.015$.

\subsection{SN-$z$ vs host-$z$}\label{subsec:SNonly}
Using the original Pantheon sample, \citet{Steinhardt2020} find a statistically significant shift in the cosmological parameters derived for the subset of SNe that have redshifts measured from the supernova (SN-$z$) and those that have host-galaxy redshifts (host-$z$). 
Here we revisit this analysis with our updated data. 

There are several important differences between \citet{Steinhardt2020} and our analysis: 1) \citet{Steinhardt2020} fit all $w$CDM parameters simultaneously, 2) our definition of the SN-$z$ sample is stricter than their not-host-$z$, in that we allow host emission-lines in SN-dominated spectra to be assigned to the host-$z$ sample, and 3) our allocation of SN-$z$ particularly for PS1MD and SDSS SNe differs from theirs.
We expect the SDSS classifications to differ because, as we addressed in Section \ref{sec:validhelio}, we updated 81 SN-$z$s to host-$z$s \citep[and further, applied the 
$+2.2\tten{-3}$ systematic offset to the remaining SN-$z$s as determined in][]{Sako2018}.
However, the reason for the PS1MD classification differences are unclear; our classifications come directly from reported redshift uncertainties (SN-$z$ have uncertainties of 0.01 and host-$z$ 0.001).

We find that restricting the data to only those SNe with host-$z$ (variation 6 in Table~\ref{tab:results} and Figure \ref{fig:H0w}) gives $\Delta H_0=0.05\pm0.20$~\kmsMpc\ (i.e.\ $0.3\sigma$) relative to the nominal Final dataset.
By contrast, when we restrict the data to the SN-$z$ sample
(of which only 17 are in the redshift range of the $H_0$ fit), we find $\Delta H_0=-2.8\pm1.3$~\kmsMpc.

These results are broadly consistent with \citet{Steinhardt2020}, who found that $\Delta H_0= 0.45\pm 0.25$~\kmsMpc\ for the host-$z$ sample and $\Delta H_0=-0.96\pm0.50$~\kmsMpc\ for the SN-$z$ sample. 
The significance of the shift in the host-$z$ sample is lower in our case, which is likely due to the greater proportion of host-$z$ redshifts in our sample and the corrections we have implemented to the SN-$z$ based on the systematic offset correction determined by \citet{Sako2018}.  
On the other hand, the $H_0$ shift we find in the SN-$z$ sample is larger than \citet{Steinhardt2020}, but our results are of similar significance (approximately $2\sigma$ in each case) since we have fewer SNe in the SN-$z$ sample. 

\begin{figure}[ht!]
    \centering
    \includegraphics[width=\textwidth]{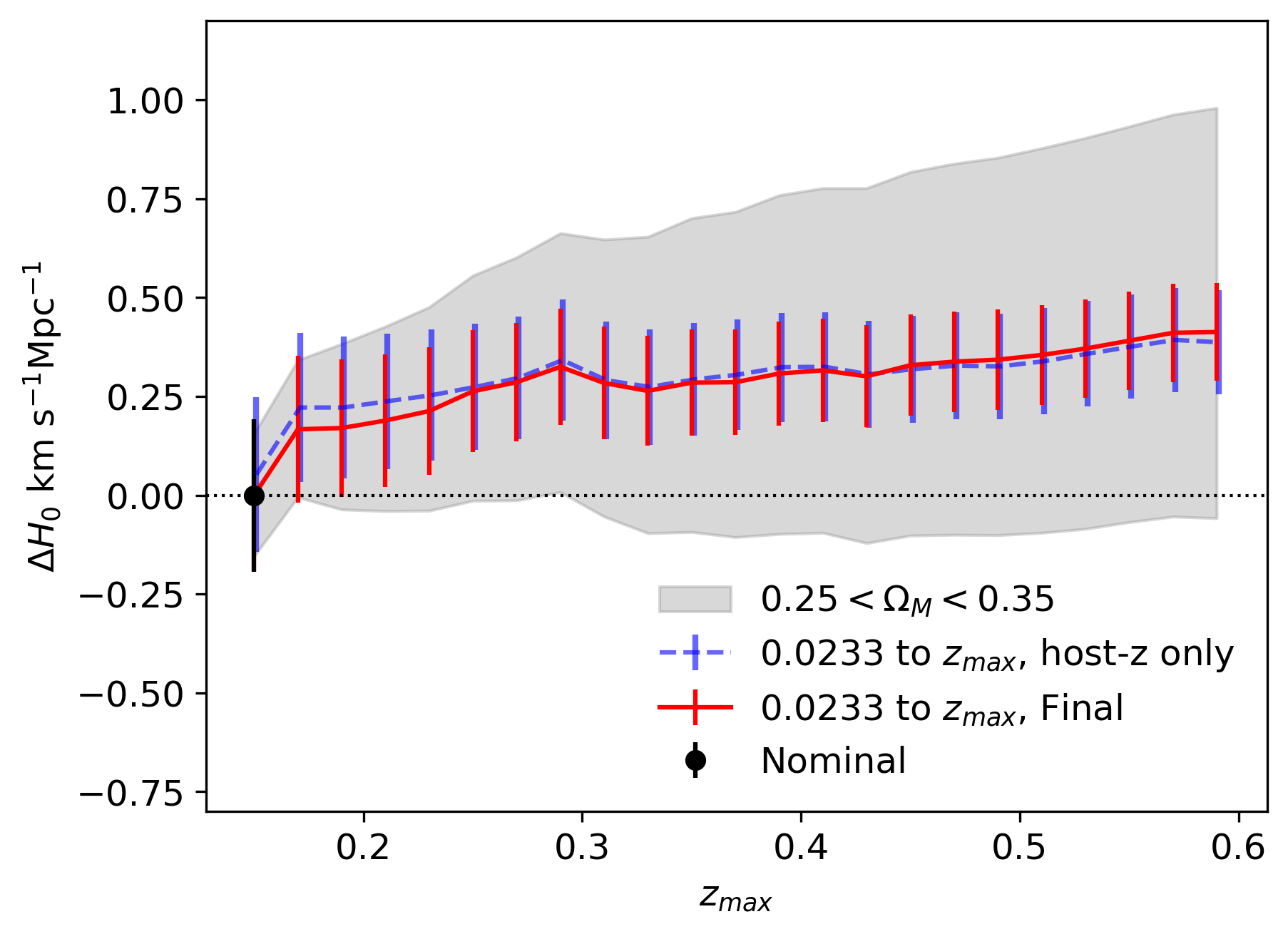}
    \caption{The dependence of $H_0$ on the redshift-range we fit over. The horizontal axis shows the maximum redshift used in the fit, keeping the minimum redshift as $\zHD{}=0.0233$. Red shows the Final sample, and blue shows the host-$z$ sample in order to demonstrate the impact of excluding SNe that only have SN-$z$. As redshift increases the cosmological model dependence becomes increasingly important so the grey shading shows the range of $H_0$ values from fitting with $0.25<\om<0.35$ in the flat-$\Lambda$CDM model ($\om=0.35$ being the lower edge). At the nominal upper redshift limit of $z_{\rm max}=0.15$ the statistical variance of the sample (black error bar) remains larger than the uncertainty due to even this quite wide range of $\om$. 
    }\label{fig:zrange}
\end{figure}

Increasing the redshift range over which we fit for $H_0$ adds some model dependence, but allows us to include more of the SN-$z$ sample. 
When we do so the deviation from the nominal dataset vanishes, with the results from SN-$z$ alone agreeing with the result from host-$z$ alone (within $1\sigma$) for all $z_{\rm max}\gtrsim 0.25$.
Figure \ref{fig:zrange} shows the impact of including or excluding SN-$z$ from our fit for $H_0$, as a function of redshift range used in the fit. 
The impact is small, with $|\Delta H_0|\lesssim 0.05$~\kmsMpc.

As expected, Figure~\ref{fig:zrange} shows that as we increase the maximum redshift in our $H_0$ fit the cosmological model-dependence becomes increasingly apparent.  
The $v_{\rm approx}(z,q_0,j_0)$ equation in Section~\ref{subsec:H0} is a good approximation to the full equation $v(z)=c\int_0^z\frac{\text{d}z}{E(z)}$ where $E(z)\equiv H(z)/H_0$, as long as $\om\sim0.3$ for the flat-$\Lambda$CDM model.
Over the nominal redshift range of $0.0233<\zHD{}<0.15$ (black point) a shift of $\Delta\om\pm0.05$ results in $\Delta H_0\mp0.2$ \kmsMpc, showing the cosmological model-dependence is still sub-dominant to the sampling uncertainties on $H_0$ (which is the purpose of using the restricted redshift range).  

For the equation of state of dark energy, the host-$z$ only case (6) shows a shift of $\Delta w = -0.022\pm0.048$.
This is the largest impact on $w$ of any variation, albeit still insignificant ($0.6\sigma$).
Using only SN-$z$ gives $\Delta w=0.25\pm0.16$. 
The reason for this shift can be seen in Figure~\ref{fig:mudiff_all}, where the SN-$z$ sample shows a systematic positive offset in $\Delta \mu$ at low redshift, but a negative offset at high redshift. 
Any shift with a redshift dependence will have an impact on $w$ whereas a constant shift would mainly affect $H_0$.  

Given these results, in Section \ref{sec:discussion_conclusion} we discuss whether SN-$z$ samples should be included in cosmological fits.

\subsection{Redshift systematics and uncertainty values}
As expected, some of the largest changes in $H_0$ and $w$ occur when we add a systematic shift to all redshifts in the sample (variations 8--11). 
However, even when we shift the redshifts systematically by their uncertainties (variations 8--9) the impact is only $|\Delta H_0|\sim 0.2\pm0.2$ \kmsMpc\ and $|\Delta w|\sim0.015\pm0.045$, smaller than the sample uncertainties in each case.  

Changing the size of the uncertainties on the redshifts has a negligible impact on the cosmological parameters (variations 12--13).  

\section{Discussion and Conclusions}\label{sec:discussion_conclusion}

Motivated in part by the fact that a systematic error in redshift could affect standard candle derivations of $H_0$, especially if it were at low redshift, we have reviewed and revised the redshifts for the Pantheon+ supernova sample. 
This includes fixing bookkeeping errors, updating heliocentric redshifts when available, adding uncertainty estimates, converting from the heliocentric to the CMB-frame redshifts without using the low-$z$ additive approximation, updating peculiar velocity estimates at all redshifts, and correcting the peculiar velocities from large-scale flow predictions at large redshifts. 

These curated redshifts are the ones that should be used in the future for all supernova cosmology analyses using these data.
They are available at the CDS VizieR database and also \data{}, and the code that generated the peculiar velocity predictions can be found at \pvcode{}.

We found that these redshift updates do not have a large impact on cosmological results.
This fortunate circumstance arises for a few reasons.
Firstly, errors at high-$z$ are relatively unimportant, because the relative uncertainty in redshift decreases as redshift increases.
That is compounded by the fact that a particular $\Delta z$ corresponds to a much smaller {\em distance} difference at high-$z$ than at low-$z$.  Furthermore the $\Delta\mu$ versus $\Delta z$ correlation we mentioned in Section~\ref{subsec:dmudz} and Figure~\ref{fig:dmudz} reduces the impact of redshift errors, especially near $z\sim0.5$.
Thus at high-$z$ even a large error in redshift gives a small error in expected magnitude.

Secondly, for the SNe at very low redshifts ($z<0.01$) that are used to calibrate the SN Ia absolute magnitude, the redshifts are {\em not} used---they are replaced by the brightness of another standard candle (TRGB or Cepheids), that acts as an anchor.

Finally, many potential systematic redshift errors (for example due to the approximate heliocentric to CMB-frame conversion) are only systematic if the sample covers a small area of the sky. 
The low-$z$ supernova sample ($z\lesssim0.15$), for which redshift errors could have a large impact, is spread across most of the sky (see Figure \ref{fig:skyplot}).
We also confirmed in Section \ref{subsec:newPVs} that the new peculiar velocities do not \textit{systematically} differ from those previously predicted.
Thus the updates to the redshifts of Pantheon+ mostly caused random shifts, predominantly via updating \zhel{} and \vp{}.
Across the whole sample the root-mean-square deviation of all the redshift changes was $\sim3\tten{-3}$, however the mean redshift change was an order of magnitude smaller ($\sim 4\tten{-4}$).  
Within the redshift range $0.0233<\zHD{}<0.15$, RMS$(\Delta\zHD{})\sim 1\tten{-3}$ and the mean redshift change was $\sim 1\tten{-4}$.
The mean change is actually almost completely frame and redshift-range independent, but is smaller for the $H_0$ sample because, unlike for the mean \zhel{} of the full sample, it does not account for repeat observations of the same SNe originally being assigned different \zhel{}.
This resulting impact on $H_0$ is consistent with that predicted by \citet[][see the green dashed line in their Figure 4]{Davis2019}.  

We compared the cosmological results with and without the sample that has only supernova redshifts in Section \ref{subsec:SNonly}).  
As in previous studies \citep[e.g.][]{Steinhardt2020}, we found differences in the results from these sub-samples, the most significant of which was a $2\sigma$ deviation in $H_0$ for the SN-$z$ sub-sample.  
While it is possible this is a statistical fluctuation with so few SNe in the SN-$z$ sample, there are reasons to expect a systematic offset from the host-$z$ sample. 
When host galaxies lack redshifts it is usually because they are faint and/or low-mass, and SNe Ia properties are correlated with their hosts' properties \citep[e.g.][]{Sullivan2006,Smith2020b,Wiseman2020}.
Therefore the SN-$z$ sample could represent a physically different subset of SNe that is not accounted for in, e.g.~SALT2 modelling.
Alternatively, a slight bias could arise if, for example, supernova spectral templates do not fully account for the blueshift due to the velocity of the visible side of the supernova's photosphere. 

Given the greater uncertainty in supernova redshifts compared to host-galaxy redshifts, and the potential for bias \citep[as seen with SDSS SNe in][]{Zheng2008, Sako2018}, one could consider removing all supernovae that lack a host redshift from cosmology samples. 
This would reduce the cosmologically useful sample size slightly, and thus sacrifice a small amount of precision for potentially greater accuracy. 
At $z<0.15$ approximately 3\% of the Type Ia supernova sample lacks host galaxy redshifts, but that proportion increases to approximately 10\% at high redshift.  
Excluding the SN-$z$ excludes a different proportion of the supernova population as a function of redshift.  
While this effect should be monitored for future cosmological studies (and should motivate further follow-up efforts to get host-$z$), we have shown here that the impact of including or excluding the SN-$z$ sample remains small relative to the uncertainties in the measurement (Figure~\ref{fig:zrange}). 

Finally, we have also made sure all supernova redshifts now have estimated uncertainties.
As noted by \citet{Steinhardt2020}, uncertainties are important because sampling from a symmetric redshift uncertainty systematically prefers a larger $\mu$ for a given redshift due to the sublinear nature of the $\mu(z)$ relation.
This would effectively reduce the gradient of the $\mu(z)$ relation at low-$z$ and cause samples with large uncertainties (such SN-$z$) to prefer a smaller $H_0$.
Furthermore, the precision of redshift indirectly affects one of the largest systematic uncertainties in SNe Ia analyses: the determination of their intrinsic scatter. 
If the precision is not correctly measured, more or less scatter will be attributed to the intrinsic variation of SN Ia distances, which could bias the modelling used to determine accurate distances. 
Additionally, the precision and accuracy of redshifts must be known accurately when using SNe Ia to measure growth-of-structure. 
In that case, instead of applying peculiar velocities to SNe from an external model, one uses SNe Ia to measure peculiar velocities. 

Most of the uncertainties we provide are based on the precision of a particular survey or sample. 
A better method for determining uncertainties would be to estimate them on a spectrum-by-spectrum basis. 
However, we tested the impact of changing the uncertainties (see Figure \ref{fig:H0w}) and it had a negligible affect on the cosmological results, so we conclude that the uncertainties we provide are sufficient for current data. 

While new surveys and datasets will continue to come online over the next decade, the sample presented here will not easily be replaced due to its utility for measuring $H_0$, which is rate-limited by the number of SNe in the very-nearby universe and will take another 30 years to re-accumulate. 
Because of that importance, in this work we endeavour to provide an updated and homogeneously treated set of supernova redshifts that we hope will be useful to the community. 

\begin{acknowledgement}
The authors thank A.~Whitford, C.~Chang, Y.~Lai, A.~Glanville and L.~Rauf for assisting in visual inspection of host galaxies and performing redshift checks in NED (Section \ref{sec:validhelio}).  We also thank C.~Howlett, M.~Colless, and M.~Hudson for useful discussions on peculiar velocities and M.~Smith for insights into SN host galaxies.
TMD is the recipient of an Australian Research Council Australian Laureate Fellowship (project number FL180100168) funded by the Australian Government. DS is supported by DOE grant DE-SC0010007, DE-SC0021962 and the David and Lucile Packard Foundation. DS is supported in part by the National Aeronautics and Space Administration under Contract No.~NNG17PX03C issued through the Roman Science Investigation Teams Programme.

This work has made use of the NASA/IPAC Extragalactic Database, which is funded by the National Aeronautics and Space Administration and operated by the California Institute of Technology.
This work was also supported by resources provided by the University of Chicago Research Computing Center, and based in part on data acquired at the Anglo-Australian Telescope. We acknowledge the traditional custodians of the land on which the AAT stands, the Gamilaraay people, and pay our respects to elders past and present. 
\end{acknowledgement}

\bibliography{references}

\clearpage
\onecolumn

\appendix
\renewcommand{\thetable}{A\arabic{table}}
\renewcommand{\theHtable}{A\arabic{table}} 
\setcounter{table}{0}

\section{Supplementary Data Tables}
\begin{table}[ht!]
\caption{Averaging of multiple SDSS redshifts. \label{tab:SDSSavgs}} 
\begin{tabular}{llcccc}
\toprule
SNID & Host & $\zhel{}$ & $\sigma_z$ & $\bar{z}$ & $\sigma_{\bar{z}}$ \\
\midrule
 & & 0.033176 & \num{1.0e-5} &  &  \\ 
\multirow{-2}{*}{2003cq} & \multirow{-2}{*}{NGC 3978} & 0.033211 & \num{1.2e-5} & \multirow{-2}{*}{0.033194} & \multirow{-2}{*}{\num{1.8e-5}} \\
\midrule 
& & 0.006352 & \num{1.2e-5} & \\ 
 & & 0.006373 & \num{1.2e-5} & \\
\multirow{-3}{*}{2003du} & \multirow{-3}{*}{UGC 9391} & 0.006406 & \num{1.3e-5} & \multirow{-3}{*}{0.006377} & \multirow{-3}{*}{\num{1.6e-5}} \\
\midrule
& &  0.016897 & \num{0.5e-5} &\\
 & & 0.016923 & \num{0.7e-5} & \\
 & & 0.016865 & \num{0.7e-5} & \\
\multirow{-4}{*}{2003Y} & \multirow{-4}{*}{IC 0522} & 0.016884 & \num{0.8e-5} &  \multirow{-4}{*}{0.016892} & \multirow{-4}{*}{\num{1.2e-5}}  \\
 \midrule
& & 0.182908 & \num{2.3e-5} &\\
 & & 0.182944 & \num{1.9e-5} & \\
\multirow{-3}{*}{1580} & \multirow{-3}{*}{WISEA J030117.99-003842.4} & 0.182945 & \num{1.9e-5} &  \multirow{-3}{*}{0.182932} & \multirow{-3}{*}{\num{1.2e-5}} \\
\midrule
& & 0.028907 & \num{1.0e-5} &\\ 
\multirow{-2}{*}{2005eq} & \multirow{-2}{*}{MCG -01-09-006} & 0.028996 & \num{1.0e-5} & \multirow{-2}{*}{0.028952} & \multirow{-2}{*}{\num{4.5e-5}} \\
\midrule
& & 0.262757 & \num{5.2e-5} &\\ 
\multirow{-2}{*}{13655} & \multirow{-2}{*}{WISEA J023605.02-005939.8} & 0.262640 & \num{3.9e-5} & \multirow{-2}{*}{0.262699} & \multirow{-2}{*}{\num{5.9e-5}}  \\
\midrule
& & 0.048475 & \num{1.2e-5} & \\ 
\multirow{-2}{*}{2006cq} & \multirow{-2}{*}{IC 4239} & 0.048409 & \num{1.2e-5} &  \multirow{-2}{*}{0.048442} & \multirow{-2}{*}{\num{3.3e-5}} \\
\midrule
& & 0.132185 & \num{3.7e-5} & \\ 
\multirow{-2}{*}{18809} & \multirow{-2}{*}{WISEA J032331.35+004002.1} & 0.132133 & \num{3.5e-5} & \multirow{-2}{*}{0.132159} & \multirow{-2}{*}{\num{2.6e-5}} \\
\midrule
& & 0.246696 & \num{2.7e-5} & \\
 & & 0.246882 & \num{2.5e-5} & \\
\multirow{-3}{*}{20039} & \multirow{-3}{*}{WISEA J003931.06+010125.2} & 0.246850 & \num{2.2e-5} & \multirow{-3}{*}{0.246809} & \multirow{-3}{*}{\num{5.7e-5}} \\
\midrule
& & 0.027821 & \num{0.6e-5} &\\ 
\multirow{-2}{*}{2007su} & \multirow{-2}{*}{SDSS J221908.85+131040.4} & 0.027842 & \num{0.6e-5} & \multirow{-2}{*}{0.027832} & \multirow{-2}{*}{\num{1.1e-5}}  \\
\midrule
& & 0.052828 & \num{4.6e-5} & \\ 
\multirow{-2}{*}{2008ac} & \multirow{-2}{*}{J115345.22+482521.0} & 0.052788 & \num{8.2e-5} & \multirow{-2}{*}{0.052808} & \multirow{-2}{*}{\num{2.0e-5}} \\
\midrule
& & 0.021054 & \num{1.0e-5} &\\
\multirow{-2}{*}{2008ds} & \multirow{-2}{*}{UGC 299}  & 0.021068 & \num{1.2e-5} &  \multirow{-2}{*}{0.021061} & \multirow{-2}{*}{\num{0.7e-5}}\\
\midrule
& & 0.020745 & \num{1.2e-5} &\\ 
\multirow{-2}{*}{2010A} & \multirow{-2}{*}{UGC 2019} & 0.020765 & \num{1.4e-5} & \multirow{-2}{*}{0.020755} & \multirow{-2}{*}{\num{1.0e-5}}  \\
\midrule
& & 0.033690 & \num{1.1e-5} &\\ 
\multirow{-2}{*}{2010ag} & \multirow{-2}{*}{UGC 10679} & 0.033739 & \num{1.2e-5} & \multirow{-2}{*}{0.033715} & \multirow{-2}{*}{\num{2.5e-5}}  \\
\midrule
& & 0.018282 & \num{2.6e-5} &\\ 
\multirow{-2}{*}{2010ai} & \multirow{-2}{*}{WISEA J125925.01+275948.2} & 0.018252 & \num{3.5e-5} & \multirow{-2}{*}{0.018267} & \multirow{-2}{*}{\num{1.5e-5}}  \\
\midrule
& & 0.089493 & \num{1.3e-5} & \\ 
\multirow{-2}{*}{590194} & \multirow{-2}{*}{WISEA J084056.86+443127.4} & 0.089532 & \num{1.3e-5} & \multirow{-2}{*}{0.089513} & \multirow{-2}{*}{\num{2.0e-5}} \\
\midrule
& & 0.016161 & \num{0.6e-5} &\\ 
\multirow{-2}{*}{2011im} & \multirow{-2}{*}{NGC 7364} & 0.016164 & \num{0.8e-5} & \multirow{-2}{*}{0.016163} & \multirow{-2}{*}{\num{0.2e-5}}  \\
\midrule
& & 0.016797 & \num{1.6e-5} &\\ 
\multirow{-2}{*}{2013gs} & \multirow{-2}{*}{UGC 5066} & 0.016827 & \num{1.7e-5} & \multirow{-2}{*}{0.016812} & \multirow{-2}{*}{\num{1.5e-5}}  \\
\bottomrule
\end{tabular}
\end{table}

\clearpage
\begin{landscape}
\begin{ThreePartTable}
\begin{TableNotes}[flushleft]
\item [a] Missing heliocentric redshift, so these cases are actually CMB-frame redshift discrepancies.
\item [b] This host was assigned by FSS because it is the closest and largest of the three nearby galaxies but we do not attempt to pick a unique host (see Figure \ref{fig:host_groups}).
\end{TableNotes}
\begin{longtable}{llS[table-format=1.7]S[table-format=1.6]S[table-format=2.1e-3]p{11cm}}

\caption{Heliocentric redshift update discrepancies $\geq$ 1\tten{-3}. Discrepancies that arise from SN redshifts are generally not included (unless they are particularly large or unusual) since they are routinely larger than 1\tten{-3}. \label{tab:update_outliers}}\\
\toprule
SNID & Host & {\zold{}} & {\znew{}} & {Difference} & Comments \\
\midrule
\endfirsthead

\caption*{Heliocentric redshift update discrepancies $\geq$ 1\tten{-3} (continued).}\\
\toprule
SNID & Host & {\zold{}} & {\znew{}} & {Difference} & Comments \\
\midrule
\endhead

\bottomrule
\endfoot

\bottomrule
\insertTableNotes
\endlastfoot

2014bj & WISEA J192240.35+435317.7 & 0.005 & 0.043 & 3.8e-2 & Original ATEL adopts a redshift of 0.045 \citep{Zhang_and_Wang2014}, subsequent classification 0.043 \citep{Balam2016}, photometric host redshift of 0.042 \citep{Yan2014} is in agreement. Old redshift comes from \citet{Stahl2019}, which cites CBET 3893 \citep{Yuk2014}. The smaller redshift is a valid but less likely fit to the SN spectrum. \\
2009dc\tnote{a} & UGC 10064 & 0 & 0.021787 & 2.2e-2 & Only affects CSP DR2 redshift, and was corrected in DR3. \\ 
14782 & SDSS J205656.18-001645.0 & 0.179 & 0.1604 & -1.9e-2 & New redshift is spectroscopic host redshift, and agrees with SNID redshift of 0.165 \citep{Ostman2011}. \citet{Sako2018} publishes the old redshift 0.179, which is a SN redshift. We mention this particular example of SN to host redshift because it is an extreme outlier. \\
Strolger & \dots & 1.01 & 1.027 & 1.7e-2 & Uncertain origin of \zold{}; \znew{} matches \citet{Riess2007}. \\ 
580104 & \ldots & 0.31 & 0.3232 & 1.3e-2 & \zold{} comes from PS1MD  but was measured to higher precision by DES (580104 is the same SN as 1261579). \\ 
Mcguire & \ldots & 1.37 & 1.357 & -1.3e-2 & Potential typo; \znew{} matches \citet{Riess2007}. \\
2007co & CGCG 172-029 & 0.016962 & 0.026962 & 1.0e-2 & Appears to be an error in the SOUSA record (leading 1 should be a 2). $z_{\text{new}}$ confirmed by CfA3, LOSS records and classification CBET \citep{Blondin2007a}.\\
2002hu & MCG +06-06-012 & 0.03 & 0.0367 & 6.7e-3 & Origin of \zold{} appears to be the NED reference to IAUC 8013 \citep{Matheson2002} that incorrectly states a redshift of 0.03. \znew{} is from \citet{Matheson2002}, who state the recession velocity of the host galaxy is 11000 \kms{}, which translates to $z=0.0367$. \\
2008ad & 2MASS J12493690+2819445 & 0.050 & 0.055441 & 5.4e-3 & Appears to be a NED error; classification CBET finds $z=0.054$ \citep{Yuan2008a} while NED reports $z_{\text{old}}=0.05$ for the host name WISEA J124936.88+281944.7. \znew{} from SDSS DR13. \\
2010ai & WISEA J125925.01+275948.2 & 0.0233 & 0.018267 & -5.0e-3 & SN and host are part of a cluster, possibly resulting in $z_{\text{old}}=0.0233$. Original CBET fits a redshift of 0.014 \citep{Nakano2010}, more or less consistent with SDSS DR13 measurement $z_{\text{new}}=0.018267$ of nearest galaxy. \\
Lancaster & \dots & 1.23 & 1.235 & 5.0e-3 & Possible rounding error; \znew{} matches \citet{Riess2007}. \\
2005lz & UGC 1666 & 0.039968 & 0.044341 & 4.4e-3 & \znew{} from the 2MASS redshift survey \citep[2MRS, ][]{Huchra2012}. \zold{} appears to be the original SN $z$ of 0.04 \citep{Silverman2012, Blondin2012} circularly converted from \zCMB{}.\\
2007qe & WISEA J235412.07+272431.9 & 0.019977 & 0.02396 & 4.0e-3 & \zold{} appears to be a circular conversion of the erroneous NED redshift $z=0.02$ from \citet{Garnavich2007}, who actually quote 0.024. This is consistent with the \znew{} we take from \citet{Childress2013} who perform dedicated host spectroscopy. \\
2007mm\tnote{a} & WISEA J010546.37-004533.6 & 0.068921 & 0.065349 & -3.6e-3 & \znew{} comes from SDSS DR13 while \zold{} is the approximate heliocentric correction to the original CBET $z=0.07$ \citep{Bassett2007}.\\
2006cz & MCG-01-38-002 & 0.03833 & 0.0418 & 3.5e-3 & Origin of \zold{} uncertain. \znew{} from 2dFGRS \citep{Colless2003}. \\
2007ci & NGC 3873 & 0.021275 & 0.017954 & -3.3e-3 & Origin of \zold{} uncertain, as it appears to be closer to the average of the two nearest galaxies. We take the SDSS DR 13 redshift of the the nearest galaxy, the most likely host. \\
1997dg & WISEA J234014.14+261209.8 & 0.033960 & 0.03081 & -3.2e-3 &\zold{} is a circular conversion of the original SN redshift estimation of 0.034 from IAUC 6753 \citep{Wang1997}, whereas \znew{} is from host spectroscopy \citep{Jha2006}. \\ 
Yowie & \ldots & 0.46 & 0.457 & -3.0e-3 & Potential rounding; \znew{} matches \citet{Riess2007}.\\
2005M & NGC 2930 & 0.0220 & 0.02484 & 2.8e-3 & \zold{} from the Catalogue of Optical Radial Velocities \citep[CORV, ][]{Fouque1992}, while \znew{} is from \citet{Childress2013}. \\
2005al & NGC 5304 & 0.015202 & 0.0124 & -2.8e-3 & Only affects CSP DR2 record; CSP DR3 redshift agrees with \znew{} from \citet{Childress2013}. \\
PTF09dnp & WISEA J151925.36+493004.8 & 0.04 & 0.037304 & 2.7e-3 & Origin of \zold{} is unclear as the value of 0.04 only seems to appear in SOUSA, at \url{https://archive.stsci.edu/prepds/sousa/}. \znew{} is from SDSS DR13. \\
1999ej & NGC 495 & 0.016388 & 0.013723 & -2.7e-3 & Origin of \zold{} uncertain, \znew{} from the Third Reference Catalogue of Bright Galaxies \citep[RC3,][]{deVaucoleurs1991} (which was used in the original CfA2 publication \citep{Jha2006}).\\
2006oa & WISEA J212342.91-005034.7 & 0.059936 & 0.062573 & 2.6e-3 & \znew{} is from the SDSS II SN Survey data release \citep{Sako2018}. \zold{} appears to be the original SN $z$ of 0.06 \citep{Bassett2006} circularly converted from \zCMB{}.\\
2007cv & IC 2597 & 0.007562 & 0.009974 & 2.4e-3 & \zold{} is from \citet{Bosma_and_Freeman1993}, which disagrees with all other measurements in NED. \znew{} from 6dF DR3 \citep{Jones2009}. \\
2017erp\tnote{a} & NGC 5861 & 0.0045 & 0.006904 & 2.4e-3 & Only affects FSS record. Uncertain origin of \zold{}. \znew{} from 6dF DR3. \\
ASASSN-14lw & WISEA J010647.87-465901.4 & 0.023 & 0.0209 & -2.1e-3 & \zold{} is potentially the average redshift of the galaxy cluster around where ASASSN-14lw occurred according to ATEL 6809 \citep{Kiyota2014}. However, \znew{} is the spectroscopic host redshift measured by CSP-II \citep{Phillips2019}.\\
2007nq & UGC 595 & 0.043523 & 0.04521 & 1.7e-3 & Origin of \zold{} uncertain, \znew{} from \citet{Childress2013}. \\
1993ae & IC 126 & 0.017932 & 0.019667 & 1.7e-3 & \zold{} appears to be a circular conversion of the \zCMB{} from \citet{Jha2007}, while \znew{} is from 6dF DR3.\\
2008fr & LEDA 5069093 & 0.040656 & 0.039 & -1.7e-3 & \zold{} appears to come from the redshift 0.0407 quoted by SIMBAD, which is the least reliable SNID redshift from \citet[][Table 7]{Silverman2012}. We take \znew{} to be the original CBET \citep{Yuan2008b} which agrees with \citet{Silverman2012} Table~1. \\
2002fk & NGC 1309 & 0.0055856 & 0.007185 & 1.6e-3 & Origin of \zold{} uncertain. \znew{} taken from 6dF DR3, and agrees with SN 2012Z also hosted by NGC 1309.\\
1999aa & NGC 2595 & 0.016019 & 0.014422 & -1.6e-3 & \zold{} appears to be a double heliocentric correction. \znew{} is from SDSS DR13. \\
2019np\tnote{a} & NGC 3254 & 0.004 & 0.005502 & 1.5e-3 & Origin of \zold{} unclear. \znew{} comes from \citet{Springob2005}.\\ 
2008L & NGC 1259 & 0.017845 & 0.01928 & 1.4e-3 & \zold{} appears to be roughly the average redshift of the Perseus cluster. 
We take \znew{} to be that of the host galaxy only \citep{Jorgensen2018}. \\
2007ux & 2MASX J10091969+1459268 & 0.029324 & 0.030699 & 1.4e-3 & Origin of \zold{} uncertain, \znew{} from SDSS DR13. \\
2010cr & NGC 5177 & 0.022925 & 0.021551 & -1.4e-3 & Origin of \zold{} uncertain, \znew{} from SDSS DR13. \\
2015N & UGC 11797 & 0.013930 & 0.012499 & -1.4e-3 & \zold{} is from the Updated Zwicky Catalogue \citep[UZC, ][]{Falco1999}, while \znew{} is from 2MRS. The redshift of UGC 11797 is contentious, possibly due to interaction with UGC 11798. \\
PSNJ1628383 & NGC 6166\tnote{b} & 0.031188 & 0.029831 & -1.4e-3 & \zold{} from FSS, but \znew{} is the average of the three nearest galaxies since a unique host cannot be determined (Figure \ref{fig:host_groups}).\\
2006bz & IC 4042A & 0.0268 & 0.028115 & 1.3e-3 & \zold{} from CORV while \znew{} comes from SDSS DR6. \\
2006kf & UGC 2829 & 0.0212992 & 0.020037 & -1.3e-3 & \znew{} from 21 cm \Hi\ emission \citep{Springob2005}. \zold{} appears to be original CfA Redshift Survey $z$ of 0.021301 \citep{Huchra1999} circularly converted from \zCMB{}.\\
2008fp & ESO 428-G014 & 0.006966 & 0.005664 & -1.3e-3 & Only affects CSP DR2 record; \zold{} appears to be a double heliocentric correction. \znew{} comes from host spectroscopy \citep{Wegner2003}, and is consistent with SN spectrum \citep{Wang2008} and CSP DR3. \\
2004gc\tnote{a} & ARP 327 NED04 & 0.030715 & 0.03196 & 1.2e-3 & Only affects CSP DR2 record. Origin of \zold{} uncertain. \znew{} from \citet{Childress2013} and consistent with CSP DR3. \\
2006qo & UGC 4133 & 0.028521 & 0.029704 & 1.2e-3 & Origin of \zold{} uncertain. \znew{} from 21 cm \Hi\ emission \citep{Theureau1998}.\\
1996C & MCG+08-25-047 & 0.027016 & 0.028235 & 1.2e-3 & \zold{} appears to be a circular conversion of the original redshift \citep[weak H$\alpha$ emission,][]{Mueller1996}, while \znew{} comes from SDSS DR13. \\
2006lu & WISEA J091517.24-253600.6 & 0.054458 & 0.0534 & -1.1e-3 & Only affects CSP DR2 record; \zold{} appears to be the CMB-frame redshift. \znew{} is from \citet{Folatelli2013}, and is the same redshift quoted by CSP DR3. \\
2000B & NGC 2320 & 0.020196 & 0.019141 & -1.1e-3 & Origin of \zold{} uncertain, \znew{} from host spectroscopy \citep{vandenBosch2015}.\\
2012ht\tnote{a} & NGC 3447A & 0.003599 & 0.004646 & 1.1e-3 & This was a case of the heliocentric-frame redshift being used as the CMB-frame redshift. \znew{} comes from SDSS DR13. \\ 
2001fh & WISEA J212042.46+442359.3 & 0.0123 & 0.013303 & 1.0e-3 & Origin of \zold{} uncertain. \znew{} from \citet{vandenBosch2015}. \\
160099 & WISEA J141838.64+541054.8 & 0.100 & 0.101018 & 1.0e-3 & \znew{} is from SDSS DR 13.\\
2007cq & WISEA J221440.71+050442.3 & 0.025 & 0.02604 & 1.0e-3 & Origin of \zold{} (SOUSA record) uncertain as it does not match the classification CBET \citep{Filippenko2007}. \znew{} is from \citet{Childress2013}. \\
2007aj & SDSS J124754.53+540038.5 & 0.031019 & 0.30 & -1.0e-3 & Origin of \zold{} uncertain, and has been replaced by the original SN $z$ \citep{Quimby2007}. \\
Eagle & \dots & 1.02 & 1.019 & -1.0e-3 & Potential rounding; \znew{} matches \citet{Riess2007}.\\
Rakke & \dots & 0.74 & 0.739 & -1.0e-3 & Potential rounding; \znew{} matches \citet{Riess2007}.\\
\end{longtable}
\end{ThreePartTable}
\end{landscape}

\begin{table}
\begin{threeparttable}[ht!]
\caption{Supernovae previously without redshift uncertainties. The uncertainty is assigned to be 5\tten{-3} if the redshift is measured directly from a SN spectrum without host emission, and 9\tten{-5} if not.  \label{tab:missing_uncertainties} }
\begin{tabular}{llSc}
\toprule
SNID & Host &\zhel{} & Assigned Uncertainty \\
\midrule
 1992J & WISEA J100901.19-263836.3 & 0.0446 & 9\tten{-5} \\
 1992ae & WISEA J212817.17-613302.7 & 0.0752 & 9\tten{-5} \\
 1992aq & APMUKS(BJ) B230147.96-373644.6 & 0.101 & 9\tten{-5} \\
 1993B & WISEA J103451.50-342635.8 & 0.0696 & 9\tten{-5} \\
 1993ac & CGCG 307-023 & 0.04937 & 9\tten{-5} \\
 1995bd & UGC 03151 & 0.01539 & 9\tten{-5} \\
 1996bl & WISEA J003618.14+112335.3 & 0.036 & 9\tten{-5} \\
 2001az & UGC 10483 & 0.040695 & 9\tten{-5} \\
 2002hu\tnote{a} & MCG +06-06-012 & 0.0367 & 1.0\tten{-3} \\
 2002kf & CGCG 233-023 & 0.01930 & 9\tten{-5} \\
 2003hu & 2MASX J19113272+7753382 & 0.075 & 9\tten{-5} \\
 2006an & SDSS J121438.73+121347.7 & 0.064 & 9\tten{-5} \\
 2006is & WISEA J051734.55-234659.7 & 0.0314 & 9\tten{-5} \\
 2006lu & WISEA J091517.24-253600.6 & 0.0534 & 9\tten{-5} \\
 2006td & KUG 0155+361 & 0.015881 & 9\tten{-5} \\
 2008L & NGC 1259 & 0.01928 & 9\tten{-5}\\
 2008by & WISEA J120520.83+405645.9 & 0.045 & 9\tten{-5} \\
 2008cf & LEDA 766647 & 0.04603 & 9\tten{-5}\\
 2008fk & WISEA J023405.17+012340.2 & 0.072 & 9\tten{-5} \\
 2008gb & UGC 02427 & 0.037 & 9\tten{-5} \\
 ASASSN-16es & SDSS J115054.45+021828.1 & 0.02850\tnote{b} & 9\tten{-5}\\
 ASASSN-16hh & MCG +03-06-031 & 0.03026\tnote{b} & 9\tten{-5}\\
 ASASSN-16lc & WISEA J192901.71-515812.6 & 0.02033\tnote{b} & 9\tten{-5}\\
 2017gup & WISEA J032934.19+105825.5 & 0.02316\tnote{b} & 9\tten{-5}\\
 2017hoq & WISEA J051920.10-173647.6 & 0.02341\tnote{b} & 9\tten{-5}\\
 2018enc & WISEA J151928.86-095256.6 & 0.02389\tnote{b} & 9\tten{-5}\\
 2018fop & WISEA J011517.81-065130.5 & 0.02121\tnote{b} & 9\tten{-5}\\
 2018jjd & GALEXASC J042420.17-315913.5 & 0.02560\tnote{b} & 9\tten{-5}\\
 ASASSN-18da & WISEA J032916.56-235839.3 & 0.02200\tnote{b} & 9\tten{-5}\\
 ASASSN-18iu & WISEA J175740.54+500200.6 & 0.02230\tnote{b} & 9\tten{-5}\\
\midrule
 1996ab & Anonymous & 0.123 & 5\tten{-3} \\
 2006mp & UGC 10754 NOTES01 & 0.023 & 5\tten{-3} \\
 2007aj & SDSS J124754.53+540038.5 & 0.030 & 5\tten{-3} \\
 2007kf & WISEA J173130.93+691844.3 & 0.0467 & 5\tten{-3} \\
 2007kg & Anonymous & 0.0067 & 5\tten{-3} \\
 2007kh & SDSS J031512.10+431012.9 & 0.050 & 5\tten{-3} \\
 2010hs & WISEA J022537.66+244557.l7 & 0.076 & 5\tten{-3} \\
 LSQ13crf & WISEA J031050.33+012519.7 & 0.060 & 5\tten{-3} \\
 2014bj & WISEA J192240.35+435317.7 & 0.043 & 5\tten{-3} \\
 2017dws & WISEA J154014.24+112040.8 & 0.082 & 5\tten{-3} \\
 2017hbi & WISEA J023232.07+352854.8 & 0.040 & 5\tten{-3} \\
\bottomrule
\end{tabular}
\begin{tablenotes}[flushleft]
\item [a] 2002hu is the only exception, which is a host redshift we have inflated the uncertainty on due to particularly ambiguous reporting of redshifts.
\item [b] Redshift independently measured by \citet{Chen2020}, but redshift uncertainties have not been released at the time of writing.
\end{tablenotes}
\end{threeparttable}
\end{table}

\newpage
\clearpage
\section{Converting real-space velocities to redshift-space velocities}\label{sec:vpecgrids}
Here we detail how we transfer the 2M$++$ velocity reconstruction from real space to redshift space.  The purpose of this transformation is so that we do not need to convert redshifts to distances in order to estimate their peculiar velocities.
\subsection{Scaling the 2M$++$ real-space grid}
As in Equation \ref{eq:vp_beta_Vext}, the peculiar velocity grid must be scaled from the normalised reconstruction to best match observed peculiar velocities from the Tully-Fisher and Fundamental Plane relations.
Each real-space grid point is scaled by $\beta$ and adjusted by \Vext{}:
\begin{equation}
    \bm{v}_{\text{p}} = \beta \bm{v}_{\text{p,recon.}} + \Vext{},
\end{equation}
where $\bm{v}_{\text{p}} = (v_X, v_Y, v_Z)$ 
are the new scaled velocities at real-space grid points $\bm{r}=(X, Y, Z)$. 
The projected line of sight velocity for each grid point is
\begin{equation}
    v_{\text{proj.}} = \bm{v}_{\text{p}}\cdot\hat{\bm{r}}
\end{equation}

\subsection{Converting to supergalactic coordinates in real-space}
The 2M++ real-space grid is in galactic Cartesian coordinates, so we transform to supergalactic Cartesian coordinates by rotating via
\begin{equation}
    R = \begin{pmatrix}
    -\sin(l) & \;\;\,\cos(l) & \;\;\, 0 \\
    -\sin(b)\cos(l) & -\sin(b)\sin(l) & \;\;\,\cos(b) \\
    \;\;\,\cos(b)\cos(l) & \;\;\,\cos(b)\sin(l) & -\sin(b)
    \end{pmatrix},
\end{equation}
so that the positive $z$-direction (now $SGZ$) is in the direction of the supergalactic north pole, $(l,b)=(47.37\dg,6.32\dg)$ \citep{Lahav2000}. 
With $\bm{r}_{\text{SG}}=(SGX,SGY,SGZ)$ and $\bm{v}_{\text{SG}}=(v_{SGX},v_{SGY},v_{SGZ})$,
\begin{equation}
    \bm{r}_{\text{SG}}^{\text{T}} = R\bm{r}^{\text{T}},
\end{equation}
and
\begin{equation}
    \bm{v}_{\text{SG}}^{\text{T}} = R\bm{v}_{\text{p}}^{\text{T}}.
\end{equation}
The projected line of sight velocity using supergalactic coordinates in real-space is
\begin{equation}
    v_{\text{SG,proj.}} = \bm{v}_{\text{SG}}\cdot\hat{\bm{r}}_{\text{SG}}
\end{equation}

\subsection{Converting to redshift-space}
The distance to any grid point is
\begin{equation}
    D = \sqrt{\bm{r}\cdot\bm{r}} = \sqrt{\bm{r}_{\text{SG}}\cdot\bm{r}_{\text{SG}}} =\sqrt{{SGX}^2 +{SGY}^2 +{SGZ}^2}.
\end{equation}
We then adjust the redshift of each grid point by its associated peculiar redshift via
\begin{equation}
     z = [(1+\bar{z})(1+z_{\text{p}}) -1],
\end{equation}
where $\bar{z}$ is the cosmological redshift corresponding to $D$ in \hMpc{} using $H_0=100h$~\kmsMpc\ (making it independent of $H_0$), $\Omega_m = 0.3$ (the same value used in the reconstruction process), and $z_p \approx v_{\text{SG,proj.}}/c$.
The redshift-space position vector is then the real-space position vector converted to redshift by the ratio of $z$ to $D$,
\begin{equation}
\bm{z}= \frac{z}{D} \bm{r}_{\text{SG}}.
\end{equation}
However, the grid points are now irregularly spaced.
Thus, the final step is to use inverse distance weighting to interpolate and adjust the irregularly-spaced grid to a regular grid.

\end{document}